\documentclass[prl,reprint,twocolumn,aps,amssymb,graphicx,floatfix,amsmath,superscriptaddress,showpacs,longbibliography]{revtex4-2}
\usepackage[utf8]{inputenc}
\usepackage[]{amsmath}
\usepackage{graphicx}
\usepackage[caption=false]{subfig}

\usepackage{bm}
\usepackage{amsmath,amssymb,graphicx}
\usepackage{bbold}
\usepackage[titletoc]{appendix}
\usepackage[colorlinks=true,citecolor=blue,urlcolor=blue,linkcolor=blue]{hyperref}
\usepackage{braket,titlesec,natbib}
\usepackage{dsfont}
\usepackage{comment}
\usepackage{float}
\usepackage{color}
\usepackage{hyperref}
\usepackage{graphicx}
\usepackage{soul}
\usepackage{ulem}

\definecolor{darkgreen}{rgb}{0.0, 0.42, 0.10}

\newcommand{\bse}{\begin{subequations}}
\newcommand{\ese}{\end{subequations}}

\newcommand{\beq}{\begin{equation}}
\newcommand{\eeq}{\end{equation}}
\newcommand{\bea}{\begin{eqnarray}}
\newcommand{\eea}{\end{eqnarray}}
\newcommand{\ve}{\varepsilon}

\newcommand{\rigt}{\rightarrow}

\newcommand{\bk}{{\bf k}}
\newcommand{\bp}{{\bf p}}
\newcommand{\bq}{{\bf q}}

\newcommand{\br}{{\bf r}}

\newcommand{\bA}{{\bf A}}

\newcommand{\bwt}{\begin{widetext}}
\newcommand{\ewt}{\end{widetext}}

\newcommand{\er}{\eqref}

\begin{document}
\title{Hyperbolic Spin Waves in Magnetic Polar Metals}
\author{Abhishek Kumar}
\email{abhishek.kumar@usherbrooke.ca}
\affiliation{Department of Physics and Astronomy, Rutgers University, Piscataway, New Jersey 08854, USA}
\affiliation{Département de physique and Institut quantique, Université de Sherbrooke, Sherbrooke, Québec, Canada J1K 2R1}
\author{Premala Chandra}
\affiliation{Department of Physics and Astronomy, Rutgers University, Piscataway, New Jersey 08854, USA}
\author{Pavel A. Volkov}
\email{pv184@physics.rutgers.edu}
\affiliation{Department of Physics and Astronomy, Rutgers University, Piscataway, New Jersey 08854, USA}
\affiliation{Department of Physics, Harvard University, Cambridge, Massachusetts, 02138 USA}
\affiliation{Department of Physics, University of Connecticut, Storrs, Connecticut 06269, USA}
\date{\today}

\begin{abstract}
We demonstrate the emergence of collective spin modes with hyperbolic dispersion in three-dimensional 
spin-orbit coupled 
polar metals magnetized by intrinsic ordering or applied fields. These particle-hole bound states exist for arbitrarily weak repulsive interactions; they are 
optically accessible and can be used to generate pure spin current when magnetization is tilted away from the polar axis. We suggest material hosts for these excitations and discuss their potential relevance to nanoscale spintronic and polaritonic applications.
\end{abstract}

\maketitle

{\it Introduction:} Hyperbolic metamaterials, highly anisotropic media, where the dispersion of photons changes sign as a function of orientation, have emerged as promising hosts for many optical applications including low loss waveguiding and sub-wavelength imaging \cite{kivshar:2013,zubin:2012, kildishev:2015, zubin:2014, hong:2020,dai:2015, peining:2015,pablo:2022}.  
Hyperbolic collective light-matter excitations including exciton-polaritons and plasmon-polaritons have also been studied \cite{sedov:2015,magdalena:2016,antonio:2018,shao2022infrared,ruta2023hyperbolic}. 
Despite the progress in the field, no platform hosting optically accessible hyperbolic spin waves has been proposed; they would present new ways to scale down spintronic circuits \cite{sarma:rmp} and magnonic systems \cite{chumak2019fundamentals,Barman_2021,flebus2023}, integrating them with optics.

\begin{figure}[tbp]
		\centering
	\includegraphics[width=\linewidth]{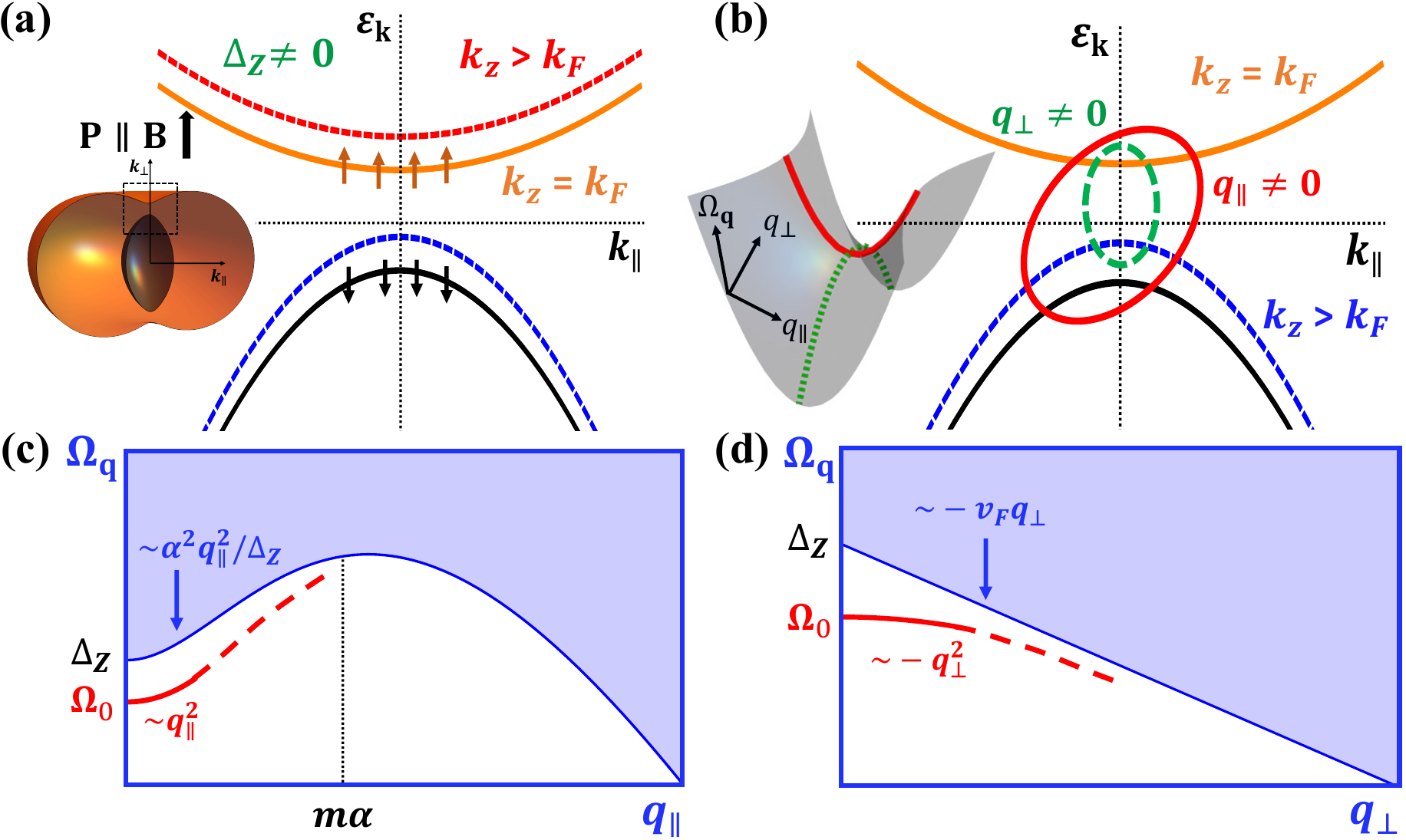}
		\caption{Schematics of (a) the single electron dispersion and (b) the particle-hole bound state near the poles of the Fermi surface (a, inset); the dispersion along $k_z$ only shifts the overall dispersion without changing its shape, implying two-dimensional kinematics for the electrons. (c) and (d) show the dispersion of particle hole excitation (c) perpendicular and (d) parallel to the polar axis. The collective mode energy is equal to $\Omega_0$ at $q=0$, and disperses with opposite curvatures at small $q$ (solid red lines),  suggesting hyperbolicity. The corresponding isofrequency surface is shown in the inset (b).}
		\label{fig:cart}
	\end{figure}

In this work we show that magnetized three-dimensional (3D) spin-orbit coupled polar metals host collective spin modes with hyperbolic dispersion (Fig.~\ref{fig:cart}). 
A magnetization component along the polar axis results in a gapped interband excitation spectrum, with a quasi-two-dimensional single-electron dispersion at momenta at minimal gap values (Figs.~\ref{fig:cart}(a) and (c)). This dimensional reduction leads to the existence of collective modes even for weak repulsive interactions, 
and their hyperbolic energy-momentum relation results from their single-particle dispersion anisotropy (Fig.~\ref{fig:cart}(b)).

We furthermore demonstrate that these spin waves, presented schematically (red) in Figs.~\ref{fig:cart}(c) and (d), carry spin close to $1$, which is polarized along the polar axis. 
Tilting the magnetization with respect to the polar axis gives the $q=0$ spin waves a nonzero velocity, resulting in a finite spin current.  
Finally, we show that these spin waves are optically accessible, suggesting possibilities for the optical generation of pure spin currents and polaritonic applications.

{\it Model:} We consider a spin-orbit coupled 3D polar metal with a closed Fermi surface and a single polar axis, chosen along $\hat{z}$ (Fig.~\ref{fig:cart}(a) inset). Requiring the presence of at least one mirror symmetry containing the polar axis, one drastically reduces the number of allowed terms in $\bk\cdot\bp$ expansion around the $\hat{z}$ axis to only two: $k_x \hat{\sigma}_y$ and $k_y \hat{\sigma}_x$. This corresponds to point groups $D_{2h}$ or $C_{3v}$.
A generic 3D Fermi surface would then cross the $z$ axis at $\pm k_F$ (note that this is not fulfilled for strongly quasi-2D materials with cylindrical Fermi surface). A low-energy model is obtained by expanding the dispersion near Dirac points on the Fermi surface $\bk=(0,0,\pm k_F)$.
Finally, we describe the effect of a magnetic field or a ferromagnetic order of magnetic impurities by Zeeman splitting - the leading term in the $k$ expansion. The resulting single-particle Hamiltonian is given by
\beq
    \hat{H}_0 =\pm v_Fk_\perp + \frac{k_\parallel^2}{2m}+ \alpha (k_y \hat{\sigma}_x-k_x \hat{\sigma}_y) 
    + \frac{\Delta_Z}{2} \hat{\sigma}_z,
    \label{eq:h0}
\eeq
where $k_\perp \equiv k_z$, $|\bk_\parallel|\equiv (k_x^2+k_y^2)^{1/2}$, $\hat{\sigma}_i$ are spin Pauli matrices and $\Delta_Z=g\mu_BB$ is the Zeeman energy splitting with $g$, $\mu_B$ and $B$ are Lande $g$-factor, Bohr magneton and the strength of the magnetic field, respectively. The single-particle Hamiltonian \er{eq:h0} is diagonal in the eigenbasis: $H_0 \to H_0^{ch} = \sum_{\bp \gamma} \ve_{\bk \gamma} a^\dagger_{\bp \gamma} a_{\bp \gamma}$, where $a_{\bp \gamma}$ is the electron annihilation operator and
\beq
\label{p-h}
\begin{split}
\ve_{\bk\gamma} = \pm v_Fk_\perp + \frac{k_\parallel^2}{2m} + \frac{\gamma}{2} \epsilon_\bk; \,\,\, \epsilon_\bk \equiv \sqrt{\Delta_Z^2+4\alpha^2k_\parallel^2},
\end{split}
\eeq
is the eigenvalue, with $\gamma=\pm$. 
A remarkable property of \er{eq:h0} is that $\epsilon_\bk$ \er{p-h}, the interband particle-hole excitation energy,  is effectively two-dimensional.
Therefore, similar to the Cooper problem \cite{cooper1956}, this reduction in dimensionality suggests the possibility of the formation of a bound state below the spin-flip particle-hole continuum (shaded region in Figs.~\ref{fig:cart}(c) and (d)) for weak interactions.

{\it Bound state at weak coupling:} To study the effects of electron-electron interactions (see the discussion of electron-soft-phonon interaction in Refs.~\cite{kumar2023, kumar_polar}), we approximate the full screened Coulomb interaction by a contact repulsive potential: 
$\hat{H}_U = (U/2V) \sum_{\bq\bp\bp'} \sum_{\sigma \sigma'} c^\dagger_{\bp+\bq, \sigma} c^\dagger_{\bp'-\bq, \sigma'} c_{\bp'\sigma'} c_{\bp\sigma}$,
where $V$ is volume and $c_{\bp \sigma}$ is the destruction operator for electron with momentum $\bp$ and spin $\sigma$.

To study the emergent particle-hole bound states and their properties, we use the random phase approximation (RPA) using equations of motion (EoM) \cite{Egri_review} approach. In the Supplementary materials (SM) \footnote{See Supplementary material for derivation and calculation of various equations presented in the Main Text.}, we show that this approach leads to same results as the diagrammatic summation of ladder diagrams for the spin-susceptibility used previously \cite{maiti_2014,maiti_2016,maslov:review}, while also allowing to access the bound state wave function. 
Another approach is phenomenological Fermi liquid theory which has been frequently considered to study spin collective modes in semiconductor heterostructures \cite{shekhter_2005, ashrafi2012, ashrafi2013, kumar2017} and graphene \cite{raines:2021, kumar:gr, raines:2022}.

Following the EoM approach \cite{Egri_review}, we consider a trial wavefunction of the particle-hole bound state with momentum $\bq$, described by the creation operator \cite{Egri_review,garate:2011}: $Q_\bq^\dagger = \sum_{\bk} C_{\bk\bq}^{+-} a_{\bk+\bq,+}^\dagger a_{\bk-}$. 
We neglect the coupling to intraband excitations - plasmons, as their energies in metals are typically large and their coupling to the interband modes vanishes linearly in $\bq$ \cite{Note1}. 
For $\Omega>0$, the coupling of $C_{\bk\bq}^{+-}$ with $C_{\bk\bq}^{-+}$ is also small, so the latter is neglected too \cite{Note1}.
The EoM for the exciton creation operator is $i \dot{Q}_\bq^\dagger (t) = [H, Q_\bq^\dagger (t)]$, and we look for stationary solutions $Q_\bq^\dagger(t) = Q_\bq^\dagger e^{-i\Omega t}$, where $\Omega$ is the spin-exciton (or bound state) frequency. 
In RPA, the operator products in the EoM are replaced by their average values in the ground state of the system \cite{Egri_review}:
$a_{\bp'\beta'}^\dagger a_{\bp \beta} \rigt \langle FG| a_{\bp'\beta'}^\dagger a_{\bp \beta} |FG \rangle = \delta_{\bp\bp'} \delta_{\beta\beta'} n_{\bp\beta}$, 
where $|FG\rangle$ refers to the ground state Fermi gas and $n_{\bp\beta} \equiv n_F(\ve_{\bp\beta})$ to the Fermi function corresponding to occupation of electronic states with momentum $\bp$ from the band $\beta$.  
The EoM results into the integral equation on $C_{\bk\bq}^{+-}$ \cite{Note1}:
\beq
\label{eq2}
\begin{split}
C_{\bk\bq}^{+-} &= N_\bq \left( \Omega_\bq - \ve_{\bk+\bq, +} + \ve_{\bk -} \right)^{-1}, \\
N_\bq &= \frac{U}{V} \sum_{\bk'} C_{\bk'\bq}^{+-} P_{\bk\bk'\bq}(n_{\bk'+\bq, +} - n_{\bk' -}),
\end{split}
\eeq
where $P_{\bk\bk'\bq}=\langle f_{\bk' -} | f_{\bk -} \rangle \langle f_{\bk+\bq, +} | f_{\bk'+\bq, +} \rangle$, with $|f_{\bp s}\rangle$ as the single-particle eigenstates of the Hamiltonian \er{eq:h0} \footnote{The interband self-energy correction is not included in the Main Text. For contact interaction, it is known to renormalize only the electron Landé-g factor appearing in the Zeeman energy ($\Delta_Z$), the details of which is delegated to the Supplementary Materials \cite{Note1}.}.

Equation~\er{eq2} can be solved for $\Omega_\bq$ only numerically because of the complicated form of $P_{\bk\bk'\bq}$. 
However, near the gapped Dirac point (DPV model) $P_{\bk\bk'\bq}$ simplifies drastically, making Eq.~\er{eq2} possible to be solved analytically. 
Indeed, near Dirac points the in-plane single-particle dispersion can be assumed weak. So, we can expand $P_{\bk\bk'\bq}$, which primarily involves $\epsilon_\bk$ \er{p-h}, assuming large $\Delta_Z$ \cite{Note1}:
$P_{\bk\bk'\bq} \approx (k_++q_+)(k_-'+q_-)/|(\bk+\bq)_\parallel(\bk'+\bq)_\parallel|$, where $p_\pm \equiv p_x+ip_y$. 
The integral equation \er{eq2}, therefore, reduces to
\beq
\label{int_1}
\begin{split}
1 &= U \sum_{\beta=\pm} \int_{-\frac{\Delta_Z}{2v_F}}^{\frac{\Delta_Z}{2v_F}} \frac{dk_\perp}{2\pi} \int \frac{d^2k_\parallel}{(2\pi)^2} \bigg[ \Omega_\bq - \Delta_Z - \beta v_Fq_z \\
&- \frac{m_++m_-}{2m_+m_-} \left( \bk_\parallel + \bq_\parallel\frac{m_-}{m_++m_-} \right)^2 - \frac{q_\parallel^2}{2(m_++m_-)} \bigg]^{-1},
\end{split}
\eeq
where $m_\pm^{-1} \equiv 2\alpha^2/\Delta_Z \pm m^{-1}$. The limits of $k_\perp$-integral are determined by the difference of Fermi functions within the Dirac point approximation.
The sum over $\beta$ takes into account the contribution from two Dirac points, located at $\pm k_F\hat{z}$ on the Fermi surface (see Fig.~\ref{fig:cart}(a)).
Since the $k_\perp$-integral \er{int_1} decouples from the $\bk_\parallel$-integral, the exciton physics near Dirac points is mainly governed by the effective two-dimensionality of the particle-hole excitation (Fig.~\ref{fig:cart}(a)), which suggests a BCS-like bound state below the spin-flip particle-hole continuum (Fig.~\ref{fig:cart}(b)). 
Writing $\Omega_\bq \approx \Delta_Z - E_B(\bq)$, where $0<E_B(\bq) \ll \Delta_Z$ is the binding energy, and then doing $\bk$-integral, we obtain $E_B(\bq) \approx \left[(E_B^0)^2 + v_F^2q_z^2\right]^{1/2} - q_\parallel^2/2(m_++m_-)$. The bound state energy (assuming small $q_z$), therefore, is
\beq
\label{dis}
\begin{split}
\Omega_\bq &\approx \Delta_Z - E_B^0 + p_1 q_\parallel^2 - p_2 q_\perp^2, \\
E_B^0 &\approx \frac{2\alpha^2\Lambda^2}{\Delta_Z} \text{Exp}\left[-\frac{8\pi^2\alpha^2v_F}{U\Delta_Z^2} \right], \,\, p_2 = \frac{v_F^2}{2E_B^0}>0, \\
p_1 &\equiv \frac{1}{2(m_++m_-)} = \frac{(4m^2\alpha^4-\Delta_Z^2)}{8m^2\alpha^2\Delta_Z},
\end{split}
\eeq
where $\Lambda$ is the high-momentum cutoff, which takes into account all the high energy contribution, including the Fermi surface. 
We note that since the $\bk_\parallel$-integral in Eq.~\er{int_1} depends on $\Lambda$, the integral equation \er{int_1} is satisfied only when the electron-electron coupling ($U$) is weak.  
Indeed, the form of $E_B^0$ suggests that the largeness coming from $\Lambda^2$ is compensated by arbitrarily weak coupling (small $U$), thus justifying $0<E_B^0\ll\Delta_Z$. This results in a bound state exponentially close to the continuum.

An interesting feature of the spin-exciton bound state \er{dis} is that its dispersion is ``hyperbolic" when $p_1>0$, or $2m\alpha^2 \gg \Delta_Z$: 
the sign of dispersion in the plane (x-y) is opposite to that out of the plane ($\hat{z}$), indicating hyperbolicity, as shown schematically (red) in Figs.~\ref{fig:cart}(c) and (d).
The hyperbolic dispersion of spin-excitons arises due to the interplay of anisotropy of the Fermi surface (evident from the eigenvalue \er{p-h}) and its concave nature near the pole (Fig.~\ref{fig:cart}(b) inset): for $p_1>0$, while the curvature of the out-of-plane dispersion for both bands is same, it is opposite in the plane near poles of the Fermi surface. This results in hyperbolic dispersion of the collective mode \er{dis}.
On the other hand, for $p_1<0$, the Fermi surface is still anisotropic, however, it is now convex which results in the mode dispersion a downward parabola \er{dis}.

We emphasize that to obtain the result \er{dis}, we assumed $E_B^0 \sim 2\alpha^2k_\parallel^2/\Delta_Z$ in Eq.~\er{int_1}. To recover the well-known large $\Delta_Z$ limit (Silin Leggett mode) \cite{silin1958, leggett1970}, one must assume $E_B^0\gg2\alpha^2k_\parallel^2/\Delta_Z$ instead (because of the largeness of $\Delta_Z$) and $k_\parallel \sim k_F$; see SM \cite{Note1} for derivations.

The appearance of large momentum cut-off ($\Lambda$) in Eq.~\er{dis} is generic for Dirac systems, hence it shows up in the DPV model.  
However, we emphasize that $\Lambda$ is not undetermined. One can assume the Fermi surface to be spherical, appropriate for a low-density semiconductor, and calculate it using diagrammatic RPA. Indeed, we find $\Lambda \sim k_F$, which is reasonable for large Fermi surface assumption; the explicit calculation gives $\Lambda \approx 2k_F e^{\alpha^2k_F^2/\Delta_Z^2}$ \cite{Note1}.
Physically, $\Lambda \sim k_F$ arises because of the change in spin polarization from the poles of the Fermi surface (fully polarized along $\hat{z}$) to its equator (in the plane), thus reducing the polarization operator.

{\it Bound State Wavefunction:} We now discuss the bound state wavefunction. Specifically, we will calculate the coefficient $C_{\bk\bq}^{+-}$, which is defined in terms of $N_\bq$ \er{eq2}. 
To obtain $N_\bq$, we use the condition $\langle FG | [ Q_\bq, Q_{\bq'}^\dagger ] | FG \rangle = \delta_{\bq\bq'}$. 
The averaging of the commutator over the ground state gives the normalization condition \cite{Note1}:
$\sum_\bk |C_{\bk\bq}^{+-}|^2 (n_{\bk-} - n_{\bk+\bq,+}) = 1$.
Using this condition in the first line of Eq.~\er{eq2}, we obtain $|N_\bq|^2 = U^{-2} \left[ \sum_\bk (n_{\bk-}-n_{\bk+\bq,+})/(\Omega_\bq - \ve_{\bk+\bq, +} + \ve_{\bk-})^2 \right]^{-1}$.
Calculating the $\bk$-integral as in Eq.~\er{int_1} and substituting it back in the first line of Eq.~\er{eq2}, we obtain \cite{Note1}
\beq
\label{wf}
\begin{split}
&|C_{\bk\bq}^{+-}|^2 \equiv \left| C_{(\pm k_{Fz}, \bk_\parallel), \bq}^{+-} \right|^2 \\
&\approx \frac{\frac{8\pi^2\alpha^2 v_F E_B^0}{V \Delta_Z^2} \left( 1 + \frac{v_F^2 q_\perp^2}{2(E_B^0)^2} \right)^{-1}}{\left[ E_B^0 + \frac{v_F^2 q_\perp^2}{2E_B^0} \pm v_F q_\perp + \frac{2\alpha^2}{\Delta_Z} \left( \bk_\parallel + \frac{\bq_\parallel}{2} \left(1+\frac{\Delta_Z}{2m\alpha^2}\right) \right)^2 \right]^2},
\end{split}
\eeq
where $\pm$ refers to two Dirac points of the Fermi surface and $E_B^0$ is given in Eq.~\er{dis}.
The normalization condition can be verified: $\sum_{\bk, \beta=\pm} |C_{(\beta k_{Fz}, \bk_\parallel), \bq}|^2 = 1$. 
The form of $|C_{\bk\bq}^{+-}|^2$ \er{wf} suggests that the exciton wavefunction is a decaying function of momenta, which indicates a localized nature of the bound state.

Using the form of $|C_{\bk\bq}|^2$ \er{wf}, we can calculate the average spin polarization in the exciton basis: $\langle\hat{\textbf{S}}\rangle \equiv \langle xc(\bq)|\hat{\textbf{S}}|xc(\bq)\rangle$, where $|xc(\bq)\rangle = Q_\bq^\dagger |FG\rangle = \sum_\bk C_{\bk\bq}^{+-} a_{\bk+\bq, +}^\dagger a_{\bk-} |FG\rangle$ is the eigenstate of spin-excitons. We find \cite{Note1}
\beq
\label{spin}
\begin{split}
\langle \hat{S}_z \rangle = 1 + \mathcal{O}(q_\parallel^2, q_z^2), \,\,\,\, \langle \hat{S}_{x(y)} \rangle = \mp \frac{q_{y(x)}}{2m\alpha} + \mathcal{O}(q_\parallel^3),
\end{split}
\eeq
where $-(+)$ sign is for $S_{x(y)}$.
We observe that $\langle\hat{S}_z\rangle \approx 1$, while $\langle\hat{S}_{x,y}\rangle$ vanishes at $q=0$. This suggests that in the weak coupling limit the exciton carries a magnetic moment of almost 1 which is oriented along $\hat{z}$.

\begin{figure}[tbp]
\centering
\includegraphics[width=\linewidth]{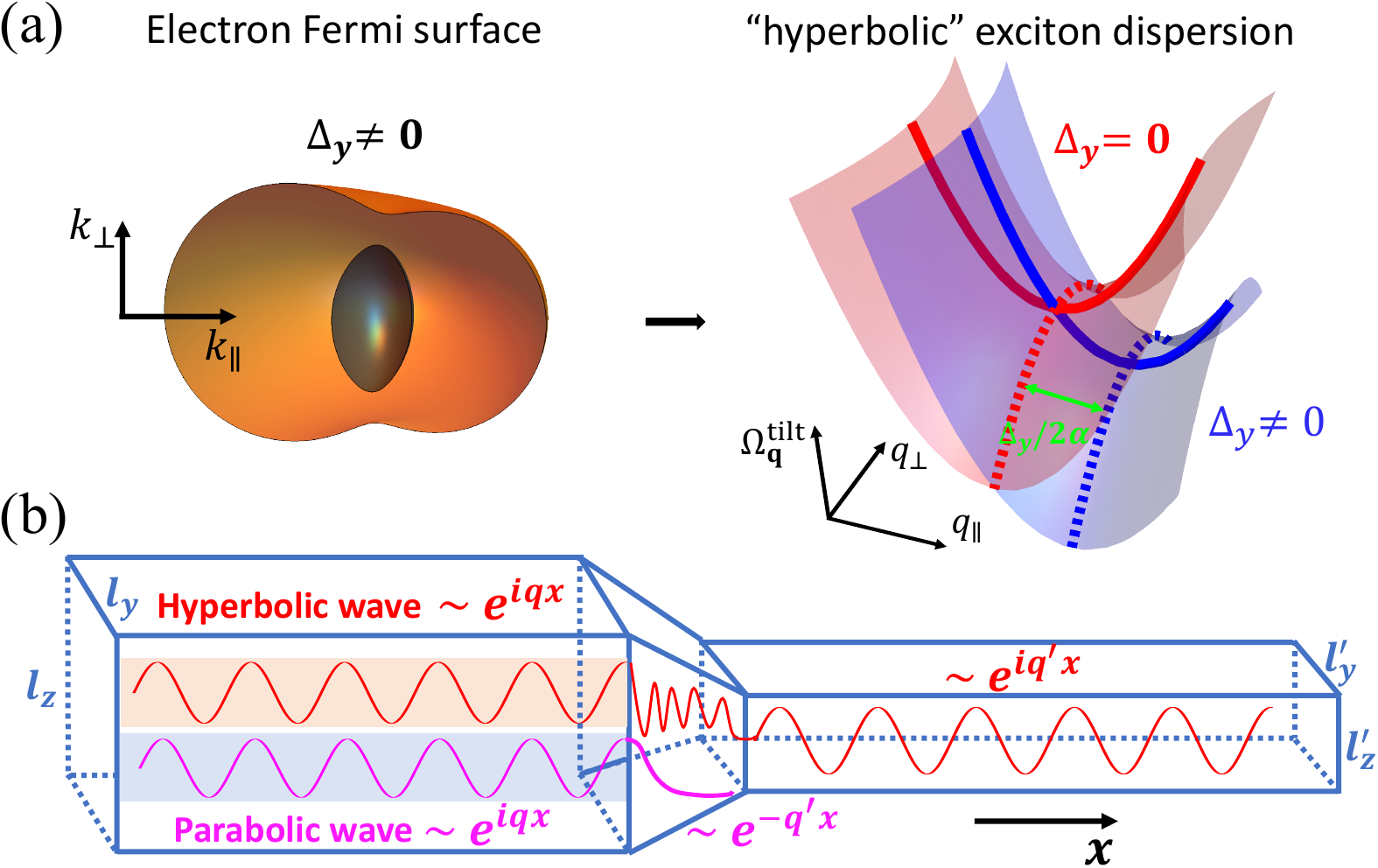}
\caption{(a) Effect of tilted magnetization ($\Delta_{y,Z} \neq 0$) with respect to the polar axis on the Fermi surface (left), and the resulting collective mode dispersion (right). (b) A schematic of a waveguide along $x$ with variable width in the transverse directions (along $\hat{y}$ and $\hat{z}$). There always exists a propagating solution inside the constriction region of the waveguide if the wave dispersion is hyperbolic. The parabolically dispersive wave is evanescent and decays away exponentially inside the constriction.}
\label{fig:cons}
\end{figure}

{\it Spin Current Generation:}
We now demonstrate that these modes can be excited optically and may carry a finite spin current in the presence of a tilted magnetic field: $\hat{H}_B = \Delta_Z/2 \hat{\sigma}_z + \Delta_y/2 \hat{\sigma}_y$, where $\Delta_Z$ and $\Delta_y$ are field components along $\hat{z}$ and $\hat{y}$ directions, respectively; see left panel of Fig.~\ref{fig:cons}(a) for the Fermi surface.  
Specifically, we show that spin-excitons propagate in the form of a wavepacket with velocity $v_{xc}$, which depends on $\Delta_y$, and their quanta carry a finite spin almost 1, polarized along the polar axis ($\hat{z}$) \er{spin}. This may lead to a pure spin-current.

We assume $\Delta_y \ll \Delta_Z$ for simplicity. 
Within the Dirac point approximation, this has negligible effect (correction would be $\mathcal{O}(\Delta_y^2)$) to the $q=0$ part of the spin-exciton frequency, $\Omega_0 \approx \Delta_Z-E_B^0$ \er{dis}, so it is dropped.
However, its dispersion modifies due to broken in-plane rotation and there appears a linear term \cite{Note1}: 
\beq
\label{dis_tilt}
\Omega_\bq^\text{tilt} = \Omega_0 + p_1(q_x^2 + q_y^2) + p_3 q_x - p_2 q_z^2; \,\,\, p_3 \equiv \Delta_y/2m\alpha,
\eeq
where $p_{1,2}>0$ \er{dis}. 
The linear term provides a boost (velocity) to the wavepacket along $\hat{x}$, which translates into a shift proportional to $\Delta_y$, as compared to that at $\Delta_y=0$, 
of the isofrequency surface of the hyperbolic mode (Fig.~\ref{fig:cons}(a) right panel). 
Finally, we find that the coefficient $|C_{\bk\bq}^{+-}|^2$ \er{wf} modifies to the form with $\bk_\parallel \to \bar{\bk}_\parallel$, where $|\bar{\bk}_\parallel| = (\bar{k}_x^2+k_y^2)^{1/2}$ and $\bar{k}_x \equiv k_x - \Delta_y/2\alpha$.

We now discuss how the spin-excitons move as a wavepacket. 
We construct a plane wave like auxiliary function $\Psi(\br, t) \equiv\sum_\bq B_\bq e^{-i\Omega_\bq^\text{tilt} t} e^{i\bq\cdot\br}$, where $\Omega_\bq^\text{tilt}$ \er{dis_tilt} is the spin-exciton frequency in the presence of $\Delta_y$, which satisfies the \textit{hyperbolic} Schrodinger equation (HSE):
\beq
\label{SE}
\begin{split}
\left( i\partial_t -\Omega_0 \right) \Psi(\br, t) = \left[ -p_1 \left( \partial_x^2 + \partial_y^2 \right) + p_2 \partial_z^2 - i p_3 \partial_x \right] \Psi(\br, t).
\end{split}
\eeq
It is to show that the exciton wavefunction takes the same form as the auxiliary function $\Psi(\br, t)$.
We define the generalized exciton state as $|\psi(t)\rangle = \sum_\bq B_\bq e^{-i\Omega_\bq^\text{tilt} t} |xc(\bq)\rangle$ and find that $\langle\psi(t)|\hat{S}_z(\br)|\psi(t)\rangle \approx |\Psi(\br, t)|^2$ at the leading order. 
Indeed, we calculate $\langle xc(\bq)|\hat{S}_z(\br)|xc(\bq')\rangle \approx e^{-i(\bq-\bq') \cdot \br} \big( 1 + \mathcal{O}\big(|\bq-\bq'|^2 \big) \big)$ \cite{Note1}, which, at leading order, suggests $\langle\psi(t)|\hat{S}_z(\br)|\psi(t)\rangle \approx \sum_{\bq, \bq'} B_\bq^* B_{\bq'} e^{-i (\bq - \bq') \cdot \br} e^{i(\Omega_\bq - \Omega_{\bq'})t} = |\Psi(\br, t)|^2$.
Hence, the spin-exciton wavefunction, which intrinsically carries a quanta of spin almost 1 and polarized along the polar axis \er{spin}, can be interpreted to be of the same form as the auxiliary function $\Psi(\br, t)$ which satisfies the HSE \er{SE}.

We now discuss the optical generation ($q\approx 0$) of spin-excitons, and the velocity with which the corresponding wavepacket moves. 
These excitons exist inside the Zeeman gap near Dirac points of the Fermi surface, as schematically shown in Figs.~\ref{fig:cart}(a) and (b). They are excited by shining the light on the sample with frequency in resonance with the optical gap. According to Fermi’s golden rule, the transition probability per unit time is defined as $\Gamma = (2\pi/\hbar) \big| \langle xc| \hat{\textbf{j}}\cdot\bA| FG \rangle \big|^2 \delta(\Omega - \Omega_0)$, where $\hat{\textbf{j}}$ is the current operator and $\bA$ is the vector potential. For our DPV model, we obtain $\Gamma$ (per unit volume) \cite{Note1}:
\beq
\label{fermi}
\begin{split}
\frac{\Gamma}{V} = \frac{e^2 \Delta_Z^2 E_B^0}{4\pi \hbar v_F} \left(\text{log}\frac{2\alpha^2\Lambda^2}{E_B^0\Delta_Z}\right)^2 (A_x^2 + A_y^2) \delta(\Omega - \Omega_0).
\end{split}
\eeq
Here $\Lambda \gg \sqrt{E_B^0\Delta_Z/2\alpha^2}$ is the same high-momentum cut off as introduced in Eq.~\er{dis}.
The spin-exciton dispersion \er{dis_tilt} suggests that at $q=0$ excitons move with velocity $v_{xc} \equiv d\Omega_\bq^\text{tilt}/dq_x|_{q_x\to 0} = p_3 = \Delta_y/2m\alpha$ along $\hat{x}$. 
So, polar metals allow for optical generation of spin-excitons which carry spin almost 1, polarized along $\hat{z}$, and move with velocity $v_{xc}=\Delta_y/2m\alpha$.  
This may lead to pure spin currents (any accompanying charge current is zero, see SM \cite{Note1}) and is the main result of this work.

We can now estimate the number of spin-excitons per unit volume ($n_{xc}$) from \er{fermi}. The exciton velocity ($v_{xc}$) is known from experiments, which, together with $n_{xc}$ and $\langle \hat{S}_z \rangle$ \er{spin}, gives the magnitude of spin-current density, defined as $j_x^{s_z} \equiv n_{xc} v_{xc} \langle \hat{S}_z \rangle$. 
Assuming finite lifetime ($\Delta^{-1}$) of excitons, in vacuum the $\Gamma/V$ near the resonance ($\Omega\approx\Omega_0\approx\Delta_Z$) can be written as (recovering correct powers of $\hbar$) $\Gamma/V \approx (2Ke^2E_B^0/\pi c\hbar^2 v_F\Delta)(P/A)(\text{log}[.])^2$ \cite{Note1}, where $K\equiv 1/4\pi\ve_0$ is a constant, with $\ve_0$ as the dielectric permittivity of free space, and $P/A$ is the power per unit area. Here, the argument of log[.] is same as that in Eq.~\er{fermi}.
For the purpose of estimation, we use this form of $\Gamma/V$, which is valid in vacuum, as a proxy to provide a ballpark for the spin-current density in a material of our interest.
The typical number of excitons per unit volume ($n_{xc}$) with power density $P/A \approx 1$ mW/mm$^2$ generated in a bulk semiconductor Cu$_2$O is $\sim 10^{15}$ cm$^{-3}$ \cite{morita:2022}.
Considering $P/A$ same as above, $E_B^0 \approx 1$ meV, $\Delta \approx 0.1$ meV, $v_F \approx 10^7$ cm/s and $\text{log}[.] \approx 10$, we get $n_{xc} \equiv \hbar\Gamma/V\Delta \approx 3\times10^{15}$ cm$^{-3}$.  
We speculate that roughly these many excitons are also optically excited in magnetic polar semiconductors, one of the examples of which is Mn-doped GeTe.
The relevant parameters of Mn-doped GeTe are: $\alpha \approx 2$ eV/\AA \, \cite{yoshimi:2018}, $m \approx 6.15 m_e$ (obtained from the measurement of Seeback coefficient \cite{zheng:2018}) and $\Delta_Z \approx 100$ meV \cite{strocov:2016}. 
Note that the carrier effective mass of parent GeTe is roughly 1.6$m_e$ \cite{zihang:2018} (density of states effective mass), which enhances to 6.15$m_e$ \cite{zheng:2018} upon Mn doping.
For an applied in-plane magnetic field, $\Delta_y \approx 1$ meV ($\ll \Delta_Z$), we get exciton velocity $v_{xc} \approx 4.7\times 10^3$ cm/s. 
Finally, the spin-current density flowing along $\hat{x}$ and polarized along $\hat{z}$ is $j_x^{s_z} \equiv n_{xc} v_{xc} \langle \hat{S}_z \rangle \approx 1.4\times 10^{19}$ cm$^{-2}$/s. 
One can verify the hyperbolicity condition, $2m\alpha^2 \gg \Delta_Z$, to be satisfied from the above mentioned values of $\alpha$, $\Delta_Z$ and $m_e$.

{\it X Waves:}
Hyperbolic dispersion of spin-excitons \er{dis_tilt} suggests the representation of spin-excitonic wave as non-spreading envelope $X$ waves. 
In general, $X$-waves are shape-preserving solutions of the non-linear HSE which has been widely studied in optical lattices \cite{stefano:2004, trull:2003, nikolaos:2009, sedov:2015, magdalena:2016, antonio:2018, huang:2020}. However, in the linear limit of the HSE, such as Eq.~\er{SE}, there exists a fundamental $X$ wave solution. 
The general $X$ wave solution of the linear HSE without the first derivative term has been discussed in Ref.~\cite{stefano:2004}, here we extend the analysis with the first derivative term \er{SE} as well, the source of which is the in-plane magnetic field ($\Delta_y$).  
The general solution is discussed in the SM \cite{Note1}, here we provide the fundamental $X$-wave solution of the linear HSE \er{SE}:
\beq
\label{x-wave}
\begin{split}
\Psi_X(\br, t) = \sqrt{p_1p_2}e^{-i\Omega_0 t}\Big[ & p_2(x-p_3t)^2 + p_2y^2 \\
&+ p_1(\sqrt{p_2}\gamma - i z)^2 \Big]^{-1/2},
\end{split}
\eeq
where, as evident, the in-plane magnetic field provides a boost ($p_3=\Delta_y/2m\alpha$ \er{dis_tilt}) to the wavepacket along $\hat{x}$ and $\gamma$ is some arbitrary width.
So we can say that spin-excitons represent localized $X$-wave like solution of the linear HSE \er{SE}, which propagates with velocity $v_{xc}=p_3=\Delta_y/2m\alpha$ along $\hat{x}$, while carrying a quanta of spin almost 1 polarized along the polar axis ($\hat{z}$). 

 

{\it Discussion:} The resemblance of spin-exciton waves with localized $X$ wave solution \er{x-wave} may have future applications in building spintronic or logic devices of smaller sizes. Indeed, like skyrmions \cite{kiselev:2011, fert:2013, finocchio:2016, kaviraj:2022}, spin-excitons could be used to transmit stored information to longer distances with lower energy consumption.
To demonstrate this, we consider an infinitely long (assumed to be along $\hat{x}$ direction) rectangular waveguide with variable geometry along transverse (assumed to be $\hat{y}$ and $\hat{z}$) directions, as shown schematically in Fig.~\ref{fig:cons}(b).
An incident wave propagating along $\hat{x}$ with hyperbolic dispersion  will transmit through the constriction (narrow region of the waveguide in Fig.~\ref{fig:cons}(b)) as it will adjust its momenta to find available states inside. On the other hand, the parabolic wave will be evanescent and decay away exponentially inside the constriction.
To understand this, let's assume that the frequency of spin-exciton wave propagating inside the waveguide is fixed and obeys the hyperbolic dispersion relation: $\Omega = q_x^2/2m_x + q_y^2/2m_y - q_z^2/2m_z$, where $q_{y, z} \equiv n_{y, z}\pi/l_{y, z}$ is quantized. 
If the incident wavevector along $\hat{x}$ is small, then inside the constriction the quantum states $n_y$ and $n_z$ can be adjusted such that we always have $q_x^2>0$. This implies propagating solution. Furthermore, if the combination $n_y^2\pi^2/2m_yl_y^2 - n_z^2\pi^2/2m_zl_z^2$ lowers the energy then $q_x$ can always be adjusted at higher values to satisfy the dispersion relation.
This does not hold for parabolic waves: the combination $n_y^2\pi^2/2m_y l_y^2 + n_z^2\pi^2/2m_z l_z^2$ always increases the energy. So, $q_x$ should be reduced. If the incident wavevector is small enough, then inside the constriction we may have the condition $q_x^2<0$, which means evanescence. This implies that the parabolic wave will decay away exponentially inside the constriction, in contrast to the hyperbolic case.
This effect may have advantages in efficiently transmitting spin current in devices of dimension as small as at nanometer (nm) scales.

In conclusion, we have proposed a mechanism for field-tunable hyperbolic spin waves and optically generated pure spin currents in three-dimensional magnetic polar metals. 
Candidate materials to test our predictions are  Mn-doped GeTe \cite{yoshimi:2018, xiaoming:2021}, $AA'$-stacked (Fe$_{0.5}$Co$_{0.5}$)$_5$GeTe$_2$ \cite{ramamoorthy:2022}, Sr$_{1-x}$Ca$_x$TiO$_{3-\delta}$ \cite{Rischau2017} with applied tilted Zeeman field, SrTiO$_3$/BaTiO$_3$/KTaO$_3$ and layered polar semiconductors such as doped AgCrSe$_2$ \cite{kim:2023}. 
The combination of strong spin polarization and hyperbolic dispersion make these spin-exciton collective modes excellent candidates for the low loss injection and transfer of spin in spintronics and magnonic circuitry.

\begin{acknowledgments}
A.K. acknowledges support from Canada First Research Excellence Fund and by the Natural Sciences and Engineering Research Council of Canada (NSERC) under Grant No. RGPIN-2019-05312. 
P.C. is supported by DOE Basic Energy Sciences Grant No. DE- SC0020353, as was A.K. during his time at Rutgers when this project was initiated. 
P.A.V. was supported by Abrahams Postdoctoral Fellowship while he was at Rutgers.
\end{acknowledgments}

\bibliography{referenceFile}

\end{document}


\title{Supplementary materials for ``Hyperbolic Spin Waves in Magnetic Polar Metals"}
\author{Abhishek Kumar}
\email{abhishek.kumar@usherbrooke.ca}
\affiliation{Department of Physics and Astronomy, Rutgers University, Piscataway, New Jersey 08854, USA}
\affiliation{Département de physique and Institut quantique, Université de Sherbrooke, Sherbrooke, Québec, Canada J1K 2R1}
\author{Premala Chandra}
\affiliation{Department of Physics and Astronomy, Rutgers University, Piscataway, New Jersey 08854, USA}
\author{Pavel A. Volkov}
\email{pv184@physics.rutgers.edu}
\affiliation{Department of Physics and Astronomy, Rutgers University, Piscataway, New Jersey 08854, USA}
\affiliation{Department of Physics, Harvard University, Cambridge, Massachusetts, 02138 USA}
\affiliation{Department of Physics, University of Connecticut, Storrs, Connecticut 06269, USA}

\date{\today}

\maketitle
\tableofcontents
In this supplementary material (SM), we provide derivation of various equations presented in the Main Text (MT). 
We provide technical details of the calculation of collective mode and its dispersion for both the Dirac-point-vicinity (DPV) model and the spherical Fermi surface (SFS) model, as discussed in the MT. For the DPV model (Sec.~\ref{model}) we use equations of motion (EoM) approach within operator RPA (random phase approximation) to calculate collective modes, while for the SFS model (Sec.~\ref{SFS}) we use standard diagrammatic RPA.

\section{Dirac-point-vicinity model}
\label{model}
We consider full Hamiltonian of the system near Dirac points of the Fermi surface as given by Eq.~(1) of the MT. The eigenvalues and eigenstates of the single-particle Hamiltonian (Eq.~(1) of the MT) are given by
\beq
\label{es}
\begin{split}
\ve_\bk^\gamma &= \pm v_F k_\perp + \frac{k_\parallel^2}{2m} + \frac{\gamma}{2} \epsilon_\bk; \\
\epsilon_\bk &\equiv \ve_+ - \ve_- = \sqrt{\Delta_Z^2 + 4\alpha^2k_\parallel^2}, \\
\ket{f_{\bk\gamma}} &=
\left( \begin{array}{c}
\frac{i \gamma \left( \gamma \Delta_Z + \epsilon_\bk \right)^{1/2}}{\sqrt{2\epsilon_\bk}} \frac{k_-}{|\bk_\parallel|} \\
\frac{\left( -\gamma \Delta_Z + \epsilon_\bk \right)^{1/2}}{\sqrt{2 \epsilon_\bk}}
\end{array} \right),
\end{split}
\eeq
where $\gamma=\pm1$ denotes the chirality of the spin-split subbands. The electron-electron interaction is treated short ranged:
\beq
\label{hub}
\hat{H}_U = \frac{U}{2V} \sum_{\bq, \bp, \bp'} \sum_{\sigma \sigma'} c^\dagger_{\bp+\bq, \sigma} c^\dagger_{\bp'-\bq, \sigma'} c_{\bp', \sigma'} c_{\bp, \sigma}.
\eeq
For the EoM approach, it is easier to work in the chiral basis. We use the transformation \cite{garate:2011}
\beq
\begin{split}
a_{\bk +}^\dagger &= \frac{1}{\sqrt{2}} \left( \sqrt{1+\frac{\Delta_Z}{\epsilon_\bk}} c_{\bk\up}^\dagger + e^{i\phi_\bk} \sqrt{1-\frac{\Delta_Z}{\epsilon_\bk}} c_{\bk\down}^\dagger \right) \\
a_{\bk -}^\dagger &= \frac{1}{\sqrt{2}} \left( \sqrt{1-\frac{\Delta_Z}{\epsilon_\bk}} c_{\bk\up}^\dagger - e^{i\phi_\bk} \sqrt{1+\frac{\Delta_Z}{\epsilon_\bk}} c_{\bk\down}^\dagger \right)
\end{split}
\eeq
and rewrite the full Hamiltonian ($\hat{H} = \hat{H}_0 + \hat{H}_U$) in the chiral basis as \cite{garate:2011, efimkin:2012}:
\beq
\label{chiral}
\begin{split}
H^{ch} = \sum_{\bp \gamma} \ve_{\bp \gamma} a^\dagger_{\bp \gamma} a_{\bp \gamma} + \frac{U}{2V} \sum_{\bq' \bp \bp'} \sum_{\gamma_1 \gamma_2} \sum_{\gamma_1' \gamma_2'} \langle f_{\bp+\bq', \gamma_1'} | f_{\bp\gamma_1} \rangle \langle f_{\bp'-\bq', \gamma_2'} | f_{\bp'\gamma_2} \rangle a^\dagger_{\bp+\bq', \gamma_1'} a^\dagger_{\bp'-\bq', \gamma_2'} a_{\bp'\gamma_2} a_{\bp\gamma_1},
\end{split}
\eeq
where $\ve_{\bk\gamma}$ and $| f_{\bp s} \rangle$ are the eigenvalue and eigenstate \er{es} of the single-particle Hamiltonian (see Eq.~(1) of the MT), respectively.

\subsection{Operator RPA: an Equation of motion approach}
We use EoM approach \cite{efimkin:2012} to calculate the collective mode frequency and the corresponding wavefunction near the Dirac points of the Fermi surface (FS). 
The creation operator ($Q_\bq^\dagger$) for the exciton-like mode with momentum $\bq$ is defined as
\beq
\label{ex-cr}
Q_\bq^\dagger = \sum_{\bk s_1 s_2} C_{\bk\bq}^{s_1 s_2} a_{\bk+\bq, s_1}^\dagger a_{\bk s_2},
\eeq
which obeys the Heisenberg equation of motion
\beq
\label{heisen}
[H^{ch}, Q_\bq^\dagger] = \Omega_\bq Q_\bq^\dagger,
\eeq
where $\Omega_\bq$ is the exciton frequency. For the case of Hubbard interaction the EoM for the exciton creation operator can be written as
\beq
\label{comm}
\begin{split}
[H^{ch}, Q_\bq^\dagger] =& \sum_{\bk s_1 s_2} C_{\bk \bq}^{s_1 s_2} (\ve_{\bk+\bq, s_1} - \ve_{\bk s_2}) a_{\bk+\bq, s_1}^\dagger a_{\bk s_2} \\
&+ \frac{U}{V} \sum_{\bk s_1 s_2} C_{\bk \bq}^{s_1 s_2} \sum_{\bp \bq'} \sum_{\gamma_1 \gamma_1' \gamma_2'} \langle f_{\bp+\bq', \gamma_1'} | f_{\bp \gamma_1} \rangle \Big( \langle f_{\bk+\bq-\bq', \gamma_2'} | f_{\bk+\bq, s_1} \rangle  a_{\bp+\bq', \gamma_1'}^\dagger a_{\bk+\bq-\bq', \gamma_2'}^\dagger a_{\bk s_2} a_{\bp \gamma_1} \\
&\hspace{8.5cm} - \langle f_{\bk s_2} | f_{\bk+\bq', \gamma_2'} \rangle a_{\bp+\bq', \gamma_1'}^\dagger a_{\bk+\bq, s_1}^\dagger a_{\bk+\bq', \gamma_2'} a_{\bp \gamma_1} \Big).
\end{split}
\eeq
The next step is to do operator RPA which reduces the product of four fermionic operators to that of two \cite{Egri_review}.
In particular, we approximate the product of four fermionic operators with a sum of two-fermion terms such that their averages over the ground state (which is a Slater determinant for the unperturbed system) coincide. In particular, we use:
\beq
a^\dagger b^\dagger c d
\to
H_{appr} = 
 a^\dagger d
\langle b^\dagger c \rangle
+
\langle a^\dagger d \rangle
 b^\dagger c
-
 a^\dagger c
\langle b^\dagger d \rangle
-
\langle a^\dagger c \rangle
 b^\dagger d
 - 
 \langle  a^\dagger d \rangle
\langle b^\dagger c \rangle
+
 \langle  a^\dagger c \rangle
\langle b^\dagger d \rangle,
\label{eq:happr}
\eeq
\\
where $a,b,c,d$ are fermionic annihilation operators. Let us illustrate that principle
\beq
\langle a^\dagger b^\dagger c d\rangle
=
\langle a^\dagger d\rangle
\langle b^\dagger c \rangle
-
\langle a^\dagger c\rangle
\langle b^\dagger d \rangle
=
\langle H_{appr} \rangle,
\eeq
assuming that the average is taken over a Slater determinant state.

For our case, the two-operator averages take the form:
\beq
\label{rpa}
a_{\bp'\gamma'}^\dagger a_{\bp \gamma} \rigt \langle 0| a_{\bp'\gamma'}^\dagger a_{\bp \gamma} |0 \rangle = \delta_{\bp\bp'} \delta_{\gamma\gamma'} n_{\bp\gamma},
\eeq
where $n_{\bp\gamma}$ is the occupation number for electronic states with momentum $\bp$ from the band $\gamma$. 
Using Eq.~\er{rpa}, the four fermionic terms in the RHS of Eq.~\er{comm} can be split into direct (Hartree) and exchange (Fock) terms:
\beq
\label{op_rpa}
\begin{split}
a_{\bp+\bq', \gamma_1'}^\dagger a_{\bk+\bq-\bq', \gamma_2'}^\dagger a_{\bk s_2} a_{\bp \gamma_1} =& \,\, a_{\bp+\bq', \gamma_1'}^\dagger a_{\bp \gamma_1} \delta_{\bq\bq'} \delta_{\gamma_2's_2} n_{\bk s_2} + a_{\bk+\bq-\bq', \gamma_2'}^\dagger a_{\bk s_2} \delta_{\bp+\bq', \bp} \delta_{\gamma_1' \gamma_1} n_{\bp\gamma_1}  \\
&- a_{\bp+\bq', \gamma_1'}^\dagger a_{\bk s_2} \delta_{\bp, \bk+\bq-\bq'} \delta_{\gamma_1 \gamma_2'} n_{\bp\gamma_1} - a_{\bk+\bq-\bq', \gamma_2'}^\dagger a_{\bp \gamma_1} \delta_{\bp+\bq', \bk} \delta_{\gamma_1' s_2} n_{\bk s_2} \\
a_{\bp+\bq', \gamma_1'}^\dagger a_{\bk+\bq, s_1}^\dagger a_{\bk+\bq' \gamma_2'} a_{\bp \gamma_1} =& \,\, a_{\bp+\bq', \gamma_1'}^\dagger a_{\bp \gamma_1} \delta_{\bq\bq'} \delta_{\gamma_2' s_1} n_{\bk+\bq, s_1} + a_{\bk+\bq, s_1}^\dagger a_{\bk+\bq' \gamma_2'} \delta_{\bp+\bq', \bp} \delta_{\gamma_1' \gamma_1} n_{\bp\gamma_1} \\
&- a_{\bp+\bq', \gamma_1'}^\dagger a_{\bk+\bq' \gamma_2'} \delta_{\bp, \bk+\bq} \delta_{\gamma_1 s_1} n_{\bk+\bq, s_1} - a_{\bk+\bq, s_1}^\dagger a_{\bp \gamma_1} \delta_{\bp\bk} \delta_{\gamma_1' \gamma_2'} n_{\bk+\bq', \gamma_2'}.
\end{split}
\eeq 
Note that the c-number parts appearing in Eq. \er{eq:happr} vanish identically here due to nonzero momentum $\bq'$. Substituting Eq.~\er{op_rpa} into Eq.~\er{comm}, and then using Eq.~\er{heisen} we get
\beq
\label{eq}
\begin{split}
\sum_{\bk s_1 s_2} C_{\bk \bq}^{s_1 s_2} (\Omega_\bq - \ve_{\bk+\bq, s_1} + \ve_{\bk s_2}) a_{\bk+\bq, s_1}^\dagger a_{\bk s_2} &= 
\frac{U}{V}\sum_{\bk s_1s_2} \langle f_{\bk+\bq,s_1} | f_{\bk s_2} \rangle a_{\bk+\bq, s_1}^\dagger a_{\bk s_2} \sum_{\bk' \gamma_1\gamma_2} C_{\bk'\bq}^{\gamma_1\gamma_2} \langle f_{\bk' \gamma_2}|f_{\bk'+\bq,\gamma_1}\rangle \\
&\hspace{6.5cm} \times \left( n_{\bk' \gamma_2} - n_{\bk'+\bq, \gamma_1} \right) \\
&+ \frac{U}{V} \sum_{\bk s_1s_2} C_{\bk\bq}^{s_1s_2} \sum_{\bk'\gamma_1\gamma_2} \langle f_{\bk'\gamma_1}|f_{\bk'\gamma_1} \rangle n_{\bk'\gamma_1} \Big( \langle f_{\bk+\bq, \gamma_2}|f_{\bk+\bq, s_1} \rangle a_{\bk+\bq, \gamma_2}^\dagger a_{\bk s_2} \\
&\hspace{6cm} - \langle f_{\bk s_2}|f_{\bk \gamma_2} \rangle a_{\bk+\bq, s_1}^\dagger a_{\bk \gamma_2} \Big) \\
&+ \frac{U}{V} \sum_{\bk s_1s_2} a_{\bk+\bq, s_1}^\dagger a_{\bk s_2} \sum_{\bk' \gamma_1\gamma_2}  C_{\bk'\bq}^{\gamma_1\gamma_2} \langle f_{\bk'\gamma_2}|f_{\bk s_2} \rangle \langle f_{\bk+\bq, s_1}|f_{\bk'+\bq, \gamma_1} \rangle  \\
&\hspace{6.5cm} \times \left( n_{\bk'+\bq, \gamma_1} - n_{\bk' \gamma_2} \right) \\
&+ \frac{U}{V} \sum_{\bk s_1s_2} C_{\bk\bq}^{s_1s_2} \sum_{\bk'\gamma_1\gamma_2} \Big( \langle f_{\bk'\gamma_2}|f_{\bk\gamma_1} \rangle \langle f_{\bk s_2}|f_{\bk' \gamma_2} \rangle a_{\bk+\bq, s_1}^\dagger a_{\bk\gamma_1} \\
&\hspace{3cm} - \langle f_{\bk'\gamma_2}|f_{\bk+\bq, s_1} \rangle \langle f_{\bk+\bq, \gamma_1}|f_{\bk', \gamma_2} \rangle a_{\bk+\bq, \gamma_1}^\dagger a_{\bk s_2} \Big) n_{\bk', \gamma_2}.
\end{split}
\eeq
The second and fourth terms in the RHS of Eq.~\er{eq} are identified as Hartree (tadpole diagram) and Fock (exchange diagram) self-energy corrections, respectively. 
Indeed, same terms are obtained if the interacting part of the Hamiltonian (second term of Eq.~\er{chiral}) is modified using the Hartree-Fock approximation: 
\beq
\delta H_{HF} = \frac{U}{V} \sum_{\bp \bp'} \sum_{\gamma_1\gamma_2' \gamma_2} \langle f_{\bp'\gamma_1}|f_{\bp'\gamma_1} \rangle \langle f_{\bp\gamma_2'}|f_{\bp\gamma_2} \rangle n_{\bp'\gamma_1} a^\dagger_{\bp\gamma_2'} a_{\bp\gamma_2}
- \frac{U}{V} \sum_{\bp \bp'} \sum_{\gamma_1\gamma_2'\gamma_2} \langle f_{\bp'\gamma_2}|f_{\bp\gamma_1} \rangle \langle f_{\bp\gamma_2'}|f_{\bp'\gamma_2} \rangle n_{\bp'\gamma_2} a^\dagger_{\bp\gamma_2'} a_{\bp\gamma_1} + const.
\eeq
Indeed, evaluating $[\delta H_{HF}, Q_\bq^\dagger]$ we obtain:
\beq
\label{self}
\begin{split}
[\delta H_{HF}, Q_\bq^\dagger] &= 
\frac{U}{V} \sum_{\bk s_1s_2} C_{\bk\bq}^{s_1s_2} \sum_{\bp \bp'} \sum_{\gamma_1\gamma_2} 
\langle f_{\bp'\gamma_1} | f_{\bp'\gamma_1} \rangle 
n_{\bp'\gamma_1}
\left( \langle f_{\bk+\bq, \gamma_2} | f_{\bk+\bq, s_1} \rangle a^\dagger_{\bk+\bq, \gamma_2} a_{\bk s_2} 
- 
\langle f_{\bk s_2} | f_{\bk\gamma_2} \rangle a^\dagger_{\bk+\bq, s_1} a_{\bk \gamma_2} \right) \\
&- \frac{U}{V} \sum_{\bk s_1s_2} C_{\bk\bq}^{s_1s_2} \sum_{\bp \bp'} \sum_{\gamma_1\gamma_2} 
\Big( \langle f_{\bp'\gamma_2} | f_{\bk+\bq,s_1} \rangle 
\langle f_{\bk+\bq, \gamma_1} | f_{\bp'\gamma_2} \rangle
a^\dagger_{\bk+\bq, \gamma_1} a_{\bk s_2} \\
&\hspace{4cm} - \langle f_{\bp'\gamma_2} | f_{\bk+\bq,\gamma_1} \rangle 
\langle f_{\bk+\bq, s_2} | f_{\bp'\gamma_2} \rangle
a^\dagger_{\bk+\bq, s_1} a_{\bk\gamma_2} \Big) n_{\bp'\gamma_2},
\end{split}
\eeq
which is equivalent to second and fourth terms of Eq.~\er{eq}. The physical meaning of these terms is a renormalization of the parameters of the single-particle Hamiltonian ($\hat{H}_0$, as defined in the first term of Eq.~\er{chiral}) \cite{Egri_review}. In the main text, we assume these correction to be already included in the parameters of $H_0$ and do not write them out explicitly.

The first and third terms of Eq. ~\er{eq} correspond to interaction energy of collective particle-hole excitations beyond Hartree-Fock. Importantly, the indices $\gamma_1,\gamma_2$ in these terms do not need to correspond to $s_1,s_2$, indicating that different types of modes (inter- and intra-band) can mix. To analyze these terms, we project the operator equations \er{eq} to the specific components, i.e., evaluate $\langle 0 |a_{\bk+\bq, s_1} 
\er{eq}
a_{\bk s_2}^\dagger|0\rangle$ which results in the following equation:
\beq
\begin{gathered}
C_{\bk \bq}^{s_1 s_2} (\Omega_\bq - \tilde{\ve}_{\bk+\bq, s_1} + \tilde{\ve}_{\bk s_2}) = T_1^{\bk\bq, s_1,s_2}+T_3^{\bk\bq, s_1,s_2},
\\
T_1^{\bk\bq, s_1,s_2} \equiv \frac{U}{V} \langle f_{\bk+\bq,s_1} | f_{\bk s_2} \rangle \sum_{\bk' \gamma_1\gamma_2} C_{\bk'\bq}^{\gamma_1\gamma_2} \langle f_{\bk'\gamma_2}|f_{\bk'+\bq,\gamma_1}\rangle \left( n_{\bk'\gamma_2} - n_{\bk'+\bq,\gamma_1} \right), \\
T_3^{\bk\bq, s_1,s_2} = \frac{U}{V} \sum_{\bk'\gamma_1\gamma_2} C_{\bk'\bq}^{\gamma_1\gamma_2} \langle f_{\bk'\gamma_2}|f_{\bk s_2} \rangle \langle f_{\bk+\bq, s_1}|f_{\bk'+\bq, \gamma_1} \rangle \left( n_{\bk'+\bq, \gamma_1} - n_{\bk'\gamma_2} \right).
\end{gathered}
\label{eq:full}
\eeq
Here we assumed that the single-particle energies have been renormalized by the Hartree-Fock correction as discussed above.

\subsection{Spin-exciton bound state}
\label{sec:spin-ex}
We now discuss bound state in the spin sector, which will now be called as ``spin-exciton". 
For the exciton case, we set $s_1=+$ and $s_2=-$, which refers to the transition between spin-split subbands and in the presence of interactions ($U$) result into a bound state in the spin sector. In Eq.~\er{eq:full}, we then identify three types of terms:
(1) $\gamma_1=\gamma_2$, (2) $\gamma_1=+,\gamma_2=-$ and (3) $\gamma_1=-,\gamma_2=+$. Explicitly separating these terms we have:
\beq
\begin{split}
&\hspace{4cm} \hat{T}_1^{\bk, +,-}+\hat{T}_3^{\bk, +,-}= (1)+(2)+(3), \\
(1)&: \frac{U}{V} \sum_{\bk',\gamma=\pm} C_{\bk'\bq}^{\gamma\gamma} \left( n_{\bk'\gamma} - n_{\bk'+\bq, \gamma} \right)
\big[ \langle f_{\bk+\bq,+}|f_{\bk-} \rangle \langle f_{\bk'\gamma}|f_{\bk'+\bq,\gamma}\rangle - \langle f_{\bk'\gamma}|f_{\bk-} \rangle \langle f_{\bk+\bq,+}|f_{\bk'+\bq,\gamma} \rangle \big] \\
(2)&: -\frac{U}{V} \sum_{\bk'} C_{\bk'\bq}^{+-} \left( n_{\bk'-} - n_{\bk'+\bq,+} \right)
\big[ \langle f_{\bk'-}|f_{\bk-} \rangle \langle f_{\bk+\bq,+}|f_{\bk'+\bq,+} \rangle - \langle f_{\bk+\bq,+}|f_{\bk-} \rangle \langle f_{\bk'-}|f_{\bk'+\bq,+}\rangle \big] \\
(3)&: \frac{U}{V} \sum_{\bk'} C_{\bk'\bq}^{-+} \left( n_{\bk'+\bq,-} - n_{\bk'+} \right)
\big[ \langle f_{\bk'+}|f_{\bk -} \rangle \langle f_{\bk+\bq,+}|f_{\bk'+\bq,-} \rangle - \langle f_{\bk+\bq,+}|f_{\bk-} \rangle \langle f_{\bk'+}|f_{\bk'+\bq,-}\rangle \big]
\end{split}
\label{contri}
\eeq
To simplify the results, we use the form of eigenstates \er{es} for $k_\parallel \ll \Delta_z/2\alpha$ (valid near gapped Dirac points $\pm k_F \hat{z}$) and corresponding relations:
\beq
\begin{gathered}
    |f_{\bk+}\rangle \approx \frac{1}{\sqrt{1+\alpha^2 k_\parallel^2/\Delta_Z^2}} 
    \begin{pmatrix}
    1 \\ -i\alpha k_+/\Delta_Z
    \end{pmatrix}, \,\,
    |f_{\bk-}\rangle \approx \frac{1}{\sqrt{1+\alpha^2 k_\parallel^2/\Delta_Z^2}} 
    \begin{pmatrix}
    -i\alpha k_-/\Delta_Z \\ 1
    \end{pmatrix}; \\
    \langle f_{\bk\pm}|f_{\bk'\pm}\rangle \approx 1 + O([k,k']^2), \,\,
    \langle f_{\bk\pm}|f_{\bk'\mp}\rangle \approx \frac{i\alpha(k_\mp-k'_\mp)}{\Delta_Z} + O([k,k']^3).
\end{gathered}
\label{eq:eigpv}
\eeq
Note that the overall phase factor, $ik_-/|\bk_\parallel|$, that one would obtain in the expression of $|f_{\bk+}\rangle$ \er{es} in the $k_\parallel \ll \Delta_Z/2\alpha$ limit is omitted for transparency. Indeed, the current form of eigenstates \er{eq:eigpv} suggests two spin flip bands in the large $\Delta_Z$ limit.
We observe that the contribution (1) in Eq.~\er{contri} vanishes at $\bq=0$ because the Fermi functions cancel each other and the leading order $q$-dependent correction is $\propto \alpha q/\Delta_Z$ at small $q$. 
Moreover, the wavefunction components $C^{\gamma\gamma}_{\bk'\bq}$ correspond to intraband excitations, where the collective modes are plasmons \cite{Egri_review}. 
To treat plasmons, we would need to take into account the long-range character of the Coulomb interaction before screening. 
However, since plasmon energies are different from exciton ones and typically large for metals, the corrections due to this coupling would scale as $U(\alpha q/\Delta_Z)^2/(\omega_{pl}-\Omega_{ex})$ and therefore can be neglected for weak coupling or small $\bq$.

Contribution (2) in Eq.~\er{contri} is the most important one, leading to the existence of a bound state. 
Assuming a weakly-coupled exciton, we write Eq.~\er{eq:full}, neglecting (3) [to be discussed below], at leading order in the relevant momenta compared to $\Delta_Z/2\alpha$:
\beq
\begin{split}
C_{\bk\bq}^{+-} &= -\frac{U}{V} \frac{\sum_{\bk'} C_{\bk'\bq}^{+-} \left( n_{\bk'-} - n_{\bk'+\bq,+} \right)}{\Omega_\bq - \tilde{\ve}_{\bk+\bq,+} + \tilde{\ve}_{\bk-}} \\
\Rigt 1 &= -\frac{U}{V} \sum_{\bk,a=\pm} \frac{\left( n_{\bk-} - n_{\bk+\bq,+} \right)}{\Omega_\bq - \tilde{\Delta}_Z -\frac{(\bk_\parallel+\bq_\parallel)^2}{2 m_+} - \frac{k_\parallel^2}{2 m_-} - a v_F q_z} \\
\Rigt 1 &= -\frac{U}{V} \sum_{\bk,a=\pm} \frac{\left( n_{\bk-} - n_{\bk+\bq,+} \right)}{\Omega_\bq - \tilde{\Delta}_Z - \frac{m_+ + m_-}{2m_+m_-} \left(\bk_\parallel + \bq_\parallel \frac{m_-}{m_+ + m_-}\right)^2 - \frac{q_\parallel^2}{2(m_+ + m_-)} - a v_F q_z},
\end{split}
\label{bs}
\eeq
where $m_\pm^{-1} = 2\alpha^2/\tilde{\Delta}_Z \pm m^{-1}$, $\tilde{\Delta}_Z$ is the self-energy renormalized Zeeman energy (to be discussed in Sec.~\ref{sec:self}) and $a=\pm1$ refers to two Dirac points ($\pm k_F \hat{z}$) of the Fermi surface.
Let us now solve the last line of Eq.~\er{bs}. 
First, near Dirac points we write Fermi functions as $n_{\bk-} \approx 1$ and $n_{\bk+}=0$.
Next, we split the $\bk$-integral into its parallel ($\bk_\parallel$) and perpendicular ($k_\perp$) components and make the substitution: $\bk_\parallel \to \bk_\parallel - \bq_\parallel (m_-/(m_+ + m_-))$. 
Finally, we write $\Omega_\bq \approx \tilde{\Delta}_Z - E_B(\bq)$, where $E_B(\bq)$ is the binding energy, and obtain
\beq
\label{spin-ex}
\begin{split}
1 &= U\int_{-\tilde{\Delta}_Z/2v_F}^{\tilde{\Delta}_Z/2v_F} \frac{dk_\perp}{2\pi} \int \frac{d^2k_\parallel}{(2\pi)^2} \left[ \frac{1}{E_B(\bq) + \frac{m_+ + m_-}{2m_+m_-} k_\parallel^2 + \frac{q_\parallel^2}{2(m_+ + m_-)} + v_F q_z} + \frac{1}{E_B(\bq) + \frac{m_+ + m_-}{2m_+m_-} k_\parallel^2 + \frac{q_\parallel^2}{2(m_+ + m_-)} - v_F q_z} \right] \\
&= \frac{U\tilde{\Delta}_Z}{4\pi^2 v_F} \int_0^\infty dk_\parallel k_\parallel \left[ \frac{1}{E_B^+(\bq) + \frac{m_++m_-}{2m_+m_-} k_\parallel^2} + \frac{1}{E_B^-(\bq) + \frac{m_++m_-}{2m_+m_-} k_\parallel^2} \right]; \,\,\,\, E_B^\pm(\bq) \equiv E_B(\bq) + \frac{q_\parallel^2}{2(m_++m_-)} \pm v_Fq_z \\
&= \frac{U\tilde{\Delta}_Z}{4\pi^2 v_F} \frac{m_+m_-}{m_++m_-} \text{ln} \left[ \left( \frac{m_++m_-}{2m_+m_-} \right)^2 \frac{\Lambda^4}{E_B^+(\bq) E_B^-(\bq)} \right], 
\end{split}
\eeq
where $\Lambda \gg \sqrt{2m_+m_- E_B^\pm(\bq)/(m_++m_-)}$ is the high momentum cutoff. Solving the above equation, we obtain for the collective mode frequency as $\Omega_\bq \approx \tilde{\Delta}_Z - E_B(\bq)$, where the binding energy $E_B(\bq)$ is obtained as
\beq
\label{be}
\begin{split}
E_B^+(\bq) E_B^-(\bq) &= \Lambda^4 \left( \frac{m_++m_-}{2m_+m_-} \right)^2 \text{Exp} \left[-\frac{4\pi^2 v_F}{U\Delta_Z} \frac{m_++m_-}{m_+m_-} \right] \\
\Rigt E_B(\bq) &= \sqrt{\Lambda^4 \left( \frac{m_++m_-}{2m_+m_-} \right)^2 \text{Exp} \left[-\frac{4\pi^2 v_F}{U\Delta_Z} \frac{m_++m_-}{m_+m_-} \right] + v_F^2 q_z^2} -\frac{q_\parallel^2}{2(m_++m_-)}.
\end{split}
\eeq
Finally, using $m_\pm^{-1} \equiv 2\alpha^2/\tilde{\Delta}_Z \pm m^{-1}$ and upon expanding for small $q_z$, we can write the collective mode frequency and its dispersion:
\beq
\label{ex}
\Omega_\bq \approx \tilde{\Delta}_Z - E_B^0 + \frac{q_\parallel^2}{2m} \frac{(4m^2\alpha^4 - \tilde{\Delta}_Z^2)}{4m\alpha^2 \tilde{\Delta}_Z} - \frac{v_F^2 q_z^2}{2E_B^0}; \,\,\,\, E_B^0 \equiv \frac{2\alpha^2\Lambda^2}{\tilde{\Delta}_Z} \text{Exp} \left[-\frac{8\pi^2\alpha^2 v_F}{U\tilde{\Delta}_Z^2} \right],
\eeq
where the form of $\tilde{\Delta}_Z$ is derived in Sec.~\ref{sec:self} (see Eq.~\er{se:rash1}) while discussing self-energy. 
As we can see, the mode dispersion is hyperbolic when $2m\alpha^2\gg \tilde{\Delta}_Z$, while it is parabolic when $2m\alpha^2\ll \tilde{\Delta}_Z$. Indeed, for the former, the dispersion in the plane is opposite to that perpendicular to the plane. On the other hand, for the latter, the dispersion is parabolically downward in all directions.


We can now discuss the contribution (3) of Eq.~\er{contri} and argue that it brings a small correction to the contribution (2).
Let us do this in the limit $q =0$ (corrections can be shown to be $O(\alpha^2 q^2/\Delta_Z^2)$) and in the vicinity of Dirac points, i.e., $n_-=1,n_+=0$. We get coupled equations:
\beq
\begin{split}
\left( \Omega - \tilde{\Delta}_Z - \frac{2\alpha^2\bk_\parallel^2}{\tilde{\Delta}_Z}\right) C_{\bk}^{+-} 
&= - \frac{U}{V} \sum_{\bk'} C_{\bk'}^{+-} + \frac{U}{V} \sum_{\bk'} C_{\bk'}^{-+} \langle f_{\bk'+}|f_{\bk-} \rangle \langle f_{\bk+}|f_{\bk'-} \rangle, \\
\left( \Omega + \tilde{\Delta}_Z +\frac{2\alpha^2\bk_\parallel^2}{\tilde{\Delta}_Z}\right) C_{\bk}^{-+} 
&= \frac{U}{V} \sum_{\bk'} C_{\bk'}^{-+} - \frac{U}{V} \sum_{\bk'} C_{\bk'}^{+-} \langle f_{\bk'-}|f_{\bk +} \rangle \langle f_{\bk-}|f_{\bk'+} \rangle.
\end{split}
\eeq
Let us treat the new terms perturbatively. In that case $\Omega\approx \tilde{\Delta}_Z > 0$ and therefore $C_{\bk}^{-+}$ can be approximated by
\beq
C_{\bk}^{-+} \approx \frac{U}{V \left(2\tilde{\Delta}_Z+\frac{2\alpha^2\bk_\parallel^2}{\tilde{\Delta}_Z}\right)}
\sum_{\bk'}
C_{\bk'}^{+-}
\langle f_{\bk'-}|f_{\bk +} \rangle \langle f_{\bk, -}|f_{\bk', +} \rangle 
+
O(U^2).
\eeq
This results in the correction to the equation for $C^{+-}_\bk$:
\beq
\label{cmp}
\begin{gathered}
\left( \Omega - \tilde{\Delta}_Z - \frac{2\alpha^2 k_\parallel^2}{\tilde{\Delta}_Z}\right)C_{\bk}^{+-} = -\frac{U}{V} \sum_{\bk'} C_{\bk'}^{+-} + \delta, \\
\delta \approx \frac{U}{V} \sum_{\bk'} \frac{U}{V} \sum_{\bk''} 
\frac{\langle f_{\bk'+}|f_{\bk -} \rangle \langle f_{\bk, +}|f_{\bk', -} \rangle 
\langle f_{\bk''-}|f_{\bk' +} \rangle \langle f_{\bk', -}|f_{\bk'', +} \rangle}{2\tilde{\Delta}_Z+\frac{2\alpha^2 k_\parallel'^2}{\tilde{\Delta}_Z}} C_{\bk''}^{+-}, \\
|\delta| < \frac{U}{V} \sum_{\bk''} \frac{U}{V} |C_{\bk''}^{+-}| \sum_{\bk'} \frac{1}{2\tilde{\Delta}_Z + \frac{2\alpha^2 k_\parallel'^2}{\tilde{\Delta}_Z}} 
\sim
\frac{U}{V}\frac{U \tilde{\Delta}_Z^2}{\alpha^2 v_F}\log \frac{\alpha\Lambda}{\tilde{\Delta}_Z} \sum_{\bk''} |C_{\bk''}^{+-}| 
\ll
\frac{U}{V} \sum_{\bk''} |C_{\bk''}^{+-}|.
\end{gathered}
\eeq
Note that the upper bound in the last line is likely an overestimate, since the matrix elements turn to zero when $\bk'=\bk$ or $\bk'' = \bk'$.
For the last inequality, we argue that $(U\tilde{\Delta}_Z^2/\alpha^2 v_F)\text{ln}(\alpha\Lambda/\tilde{\Delta}_Z) \ll 1$. Indeed, in Eq.~\er{spin-ex} the integral suggests that $(U\tilde{\Delta}_Z^2/\alpha^2 v_F)\text{ln}(\alpha\Lambda/E_B^0) \sim 1$. Since $E_B^0 \ll \tilde{\Delta}_Z$, the last inequality in Eq.~\er{cmp} is justified. 
This clearly demonstrates that $\delta$ is small compared to the other term in the RHS of the first line of Eq.~\er{cmp}. Hence, we can neglect the effects of coupling between $C^{+-}$ and $C^{-+}$. 

\subsection{Self-energy}
\label{sec:self}
In this section, we discuss self-energy which renormalizes parameters of the single-particle Hamiltonian $\hat{H}_0$. 
We consider second and fourth terms of Eq.~\er{eq}, also illustrated in Eq.~\er{self}, for the exciton case: $s_1=+$ and $s_2=-$.
The second term vanishes trivially, so it is only the fourth term which is relevant.
We rewrite the integral equation \er{eq:full} as
\beq
\label{se:gen}
\begin{split}
& C_{\bk\bq}^{+-} \left( \Omega_\bq - \ve_{\bk+\bq,+} + \ve_{\bk-} - \delta\Sigma_{\bk\bq} \right) = T_1^{\bk\bq,+,-} + T_3^{\bk\bq,+,-}, \\
\text{where} \,\,\, &\delta\Sigma_{\bk\bq} = \frac{U}{V} \sum_{\bk'} \Big[ \Big\{ \big| \langle f_{\bk'-}|f_{\bk-} \rangle \big|^2 - \big| \langle f_{\bk'-}|f_{\bk+\bq,+} \rangle \big|^2 \Big\} n_{\bk'-} + \Big\{ \big| \langle f_{\bk'+}|f_{\bk-} \rangle \big|^2 - \big| \langle f_{\bk'+}|f_{\bk+\bq,+} \rangle \big|^2 \Big\} n_{\bk'+} \Big] 
\end{split}
\eeq

\subsubsection{Zero Rashba case: Silin-Leggett mode}
Let us first discuss the limiting case: $q=0$ and $\alpha=0$ (no Rashba). This case must give Silin-Leggett mode which is nothing but the Larmor frequency $\Omega=\Delta_Z$ \cite{silin1958}. The self-energy is simply $\delta\Sigma_\bk = (U/V) \sum_{\bk'} (n_{\bk'-} - n_{\bk'+})$.
At $\alpha=0$, the Fermi surface is spherical so no Dirac point approximation can be used here.
Using $n_{\bk\pm} \equiv n_F(k^2/2m \pm \Delta_Z/2 - \mu)$, where $n_F$ is the Fermi function and $\mu$ is the chemical potential, and then doing $\bk$-integral assuming $\Delta_Z \ll \mu$, we get $\delta\Sigma_\bk \approx U\nu_F\Delta_Z/2$, where $\nu_F \equiv mk_F/\pi^2$ is the total 3D density of states. 
Hence, the self-energy correction leads to the renormalization of Zeeman energy: $\Delta_Z \to \tilde{\Delta}_Z \equiv \Delta_Z(1+U\nu_F/2)$.

We can now solve for the collective mode frequency \er{bs} and obtain: $\Omega = \tilde{\Delta}_Z(1-U\nu_F/2) = \Delta_Z(1 - U^2\nu_F^2/4)$. 
The Kohn's theorem states that the Silin-Leggett mode is a bare Larmor frequency, protected from renormalization due to interaction. 
This is indeed the case here and the quadratic $U$ correction is an artifact of not calculating the self-energy correction self-consistently.
Indeed, the exact self-consistent calculation of self-energy suggests $\tilde{\Delta}_Z = \Delta_Z/(1-U\nu_F/2)$ \cite{maiti_2016}. 
The factor in the denominator is cancelled exactly by the one appearing in the collective mode frequency.
So, the exact cancellation of self-energy and vertex corrections leads to the collective mode as Larmor frequency, $\Omega=\Delta_Z$. 
Hence, in our expression we simply ignore the $U^2$ correction.

\subsubsection{Finite Rashba case}
Let us now discuss self-energy correction for finite Rashba $(\alpha \neq 0)$. We will still assume $2\alpha k_\parallel \ll \Delta_Z$, which is valid near Dirac-points on the Fermi surface.
Moreover, to be consistent with an earlier work \cite{garate:2011}, we recover the additional phase factor, $ik_-/|\bk_\parallel|$, in the limiting form of $|f_{\bk+}\rangle$ near Dirac point. The eigenstate \er{eq:eigpv} then modifies to
\beq
\begin{gathered}
    |f_{\bk+}\rangle \approx \frac{ik_-/|\bk_\parallel|}{\sqrt{1+\alpha^2 k_\parallel^2/\Delta_Z^2}} 
    \begin{pmatrix}
    1 \\ -i\alpha k_+/\Delta_Z
    \end{pmatrix}, \,\,
    |f_{\bk-}\rangle \approx \frac{1}{\sqrt{1+\alpha^2 k_\parallel^2/\Delta_Z^2}} 
    \begin{pmatrix}
    -i\alpha k_-/\Delta_Z \\ 1
    \end{pmatrix};
\end{gathered}
\label{eigak}
\eeq
Using this, we obtain the form of $\delta\Sigma_{\bk\bq}$ \er{se:gen} as
\beq
\begin{split}
\delta\Sigma_{\bk\bq} &= \frac{U\Delta_Z^2}{2V} \left( \frac{1}{\epsilon_\bk} + \frac{1}{\epsilon_{\bk+\bq}} \right) \sum_{\bk'} \frac{n_{\bk'-} - n_{\bk'+}}{\epsilon_{\bk'}} \\
&= \frac{U\Delta_Z^2}{2} \left( \frac{1}{\epsilon_\bk} + \frac{1}{\epsilon_{\bk+\bq}} \right) \frac{\Delta_Z^2}{8\pi^2\alpha^2 v_F} \text{ln} \frac{2\alpha^2\Lambda^2}{\Delta_Z^2},
\end{split}
\label{se:rashba}
\eeq
where $\Lambda \gg \Delta_Z/\sqrt{2}\alpha$ is the same high-momentum cut off as obtained in the bound state energy \er{ex} above.
Also, the Fermi function, $n_{\bk'\pm}$, appearing above has been assumed to take the form near Dirac points: $n_{\bk'-}=1$ and $n_{\bk'+}=0$.
Combining self-energy \er{se:rashba} with the single-particle dispersion as in $(\Omega_\bq - \ve_{\bk+\bq,+} + \ve_{\bk-} - \delta\Sigma_{\bk\bq})$, we find that only $\Delta_Z$ renormalizes due to interaction:
\beq
\label{se:rash1}
\Delta_Z \to \tilde{\Delta}_Z \approx \Delta_Z \left(1+\frac{U\Delta_Z^2}{8\pi^2\alpha^2 v_F} \text{ln} \frac{2\alpha^2\Lambda^2}{\Delta_Z^2} \right)
\eeq
From the same logic as presented while arguing $C_\bk^{-+} \ll C_\bk^{+-}$ (see text below Eq.~\er{cmp}), we can say that the second term of Eq.~\er{se:rash1} is small compared to 1.

\subsection{Wavefunction of the collective mode frequency}
\label{sec:wf}
In this section we will provide technical details of the calculation of wavefunction (more specifically $C_{\bk\bq}^{+-}$) of the spin excitonic wave \er{ex}. 
Let us rewrite Eq.~\er{eq:full} for the exciton case (for brevity we simply write $\tilde{\Delta}_Z$ as $\Delta_Z$):
\beq
\label{eq1}
\begin{split}
C_{\bk\bq}^{+-} (\Omega_\bq - \ve_{\bk+\bq, +} + \ve_{\bk -}) = \frac{U}{V} \sum_{\bk'} C_{\bk'\bq}^{+-} \left( n_{\bk'+\bq,+} - n_{\bk'-} \right)
\big[ \langle f_{\bk'-}|f_{\bk-} \rangle \langle f_{\bk+\bq,+}|f_{\bk'+\bq,+} \rangle - \langle f_{\bk+\bq,+}|f_{\bk-} \rangle \langle f_{\bk'-}|f_{\bk'+\bq,+}\rangle \big]
\end{split}
\eeq
Using Eq.~\er{eigak}, the matrix element in Eq.~\er{eq1} within large-$\Delta_Z$ approximation can be written as
\beq
\label{mat1}
\begin{split}
\langle f_{\bk'-} | f_{\bk-} \rangle \langle f_{\bk+\bq,+} | f_{\bk'+\bq,+} \rangle &\approx \frac{(k_++q_+)}{|(\bk+\bq)_\parallel|} \frac{(k'_-+q_-)}{|(\bk'+\bq)_\parallel|}, \\
\langle f_{\bk+\bq,+} | f_{\bk-} \rangle \langle f_{\bk'-} | f_{\bk'+\bq,+} \rangle &\approx  \frac{\alpha^2 q_+q_-}{\Delta_Z^2} \frac{(k_++q_+)}{|(\bk+\bq)_\parallel|} \frac{(k'_-+q_-)}{|(\bk'+\bq)_\parallel|}.
\end{split}
\eeq
Using Eq.~\er{mat1}, the Eq.~\er{eq1} simplifies to
\beq
\label{wf}
C_{\bk\bq}^{+-} = U \left( 1 - \frac{\alpha^2 q_\parallel^2}{\Delta_Z^2} \right) \frac{(k_++q_+)}{|(\bk+\bq)_\parallel|}  \frac{N_\bq}{\Omega_\bq - \ve_{\bk+\bq, +} + \ve_{\bk -}},
\eeq
where
\beq
N_\bq \equiv \frac{1}{V}\sum_{\bk'} C_{\bk' \bq}^{+-} \frac{k'_-+ q_-}{|(\bk'+\bq)_\parallel|} (n_{\bk'+\bq, +} - n_{\bk' -}).
\eeq
Note that an extra $q_\parallel^2$ correction term shows up in Eq.~\er{wf} on the RHS, which is a result of interband matrix element at finite $q$ (second term of Eq.~\er{eq1}). This $q_\parallel^2$ correction should have been considered in the dispersion of spin-exciton frequency \er{ex} as well. However, we argue that this correction is very small, so ignored in Eq.~\er{ex}. Indeed, if we notice, this factor merely dresses the interaction: $U \to U(1-\alpha^2q_\parallel^2/\Delta_Z^2)$ \er{wf}. If we plug this back in Eq.~\er{ex}, it leads to a positive correction in $q_\parallel^2$ which is good for hyperbolicity of the dispersion. However, the correction is of order $\mathcal{O}\left( (E_B^0/\Delta_Z)(\alpha^2v_F/U\Delta_Z^2) \right)$, which is small since $E_B^0 \ll \Delta_Z$. So it was ignored.

We now calculate the normalization factor $N_\bq$. As discussed in the MT, $N_\bq$ can be obtained by using the condition, $\langle0| [Q_\bq, Q_{\bq'}^\dagger] |0\rangle = \delta_{\bq\bq'}$, where $Q_\bq^\dagger$ \er{ex-cr} is the general exciton creation operator. Using the commutation relation
\beq
\label{comm1}
\begin{split}
[ Q_\bq, Q_{\bq'}^\dagger ] = \sum_{\bk s_1 s_2} \bigg\{ \sum_{s_2'} a_{\bk s_2}^\dagger a_{\bk+\bq-\bq', s_2'} (C_{\bk\bq}^{s_1 s_2})^\dagger (C_{\bk+\bq-\bq', \bq'}^{s_1 s_2'}) - \sum_{s_1'} a_{\bk+\bq' s_1'}^\dagger a_{\bk+\bq, s_1} (C_{\bk\bq}^{s_1 s_2})^\dagger (C_{\bk\bq'}^{s_1' s_2}) \bigg\}.
\end{split}
\eeq
and then averaging over the ground state, we get the relation $\sum_{\bk s_1 s_2} |C_{\bk\bq}^{s_1 s_2}|^2 (n_{\bk s_2} - n_{\bk+\bq s_1}) = 1$. Choosing $\{s_1, s_2\} \equiv \{+, -\}$ and then using Eq.~\er{wf} in this, we get
\beq
\label{norm}
|N_\bq|^2 = \frac{1}{U^2 \left( 1- \alpha^2 q_\parallel^2/\Delta_Z^2 \right)^2} \left[ \sum_\bk \frac{n_{\bk-} - n_{\bk+\bq,+}}{(\Omega_\bq - \ve_{\bk+\bq, +} + \ve_{\bk -})^2} \right]^{-1},
\eeq
where we recall the form of $\ve_{\bp\gamma}$ in Eq.~\er{es}. The wavefunction finally is given by
\beq
\label{wf1}
|C_{\bk\bq}^{+-}|^2 = \frac{1}{(\Omega_\bq - \ve_{\bk+\bq, +} + \ve_{\bk -})^2} \left[ \sum_{\bk'} \frac{n_{\bk'-} - n_{\bk'+\bq,+}}{(\Omega_\bq - \ve_{\bk'+\bq, +} + \ve_{\bk' -})^2} \right]^{-1}.
\eeq
Now it is left to do $\bk'$ integral in the above equation \er{wf1}.

To begin the discussion, we first assume $\bq \parallel \hat{x}, \hat{y}$. The $\bk$-integral in Eq.~\er{wf1} can be done the same way as discussed in Eq.~\er{spin-ex}. 
Writing $\Omega_\bq \approx \Delta_Z - \sqrt{(E_B^0)^2 + v_F^2q_z^2} + q_\parallel^2/2(m_++m_-)$ (using Eqs.~\er{bs} - \er{ex}), where $m_\pm^{-1} = 2\alpha^2/\Delta_Z \pm m^{-1}$, we calculate:
\beq
\label{N_0}
\begin{split}
\sum_\bk \frac{n_{\bk-} - n_{\bk+\bq_\parallel,+}}{(\Omega_{\bq_\parallel} - \ve_{\bk+\bq_\parallel, +} + \ve_{\bk -})^2} &= \sum_\bk \frac{n_{\bk-} - n_{\bk+\bq_\parallel,+}}{\left( \Delta_Z - \sqrt{(E_B^0)^2 + v_F^2 q_z^2} + \frac{q_\parallel^2}{2(m_++m_-)} - \ve_{\bk+\bq_\parallel, +} + \ve_{\bk -} \right)^2} \\
&\approx \frac{\Delta_Z V}{2\pi v_F} \sum_{a=\pm} \int \frac{d^2k_\parallel}{(2\pi)^2} \frac{1}{\left( \sqrt{(E_B^0)^2 + v_F^2q_z^2} + av_Fq_z + \frac{m_++m_-}{2m_+m_-} \left( \bk_\parallel + \bq_\parallel \frac{m_-}{m_++m_-} \right) \right)^2} \\
&\approx \frac{\Delta_Z^2 V}{8\pi^2\alpha^2 v_F E_B^0} \left( 1+\frac{v_F^2 q_z^2}{2(E_B^0)^2} \right). \\
\text{Eq.~\er{norm}} \Rigt \,\, |N_\bq|^2 &\approx \frac{8\pi^2\alpha^2 v_F E_B^0}{V U^2\Delta_Z^2} \left( 1-\frac{\alpha^2q_\parallel^2}{\Delta_Z^2} \right)^{-1} \left( 1+\frac{v_F^2 q_z^2}{2(E_B^0)^2} \right)^{-1}.
\end{split}
\eeq
Substituting \er{N_0} into \er{wf1}, we get
\beq
\begin{split}
\label{wf_in-plane}
|C_{a\bk \bq_\parallel}^{+-}|^2 &\approx \frac{8\pi^2\alpha^2 v_F E_B^0}{\Delta_Z^2 V} \left( 1+\frac{v_F^2q_z^2}{2(E_B^0)^2}  \right)^{-1} \frac{1}{\left( \Omega_\bq - \Delta_Z - av_Fq_z - \frac{(\bk_\parallel + \bq_\parallel)^2}{2m_+} - \frac{k_\parallel^2}{2m_-} \right)^2} \\
&\approx \frac{8\pi^2\alpha^2 v_F E_B^0}{\Delta_Z^2 V} \left( 1+\frac{v_F^2q_z^2}{2(E_B^0)^2}  \right)^{-1} \left( \sqrt{(E_B^0)^2 + v_F^2q_z^2} + av_Fq_z + \frac{m_++m_-}{2m_+m_-} \left( \bk_\parallel + \bq_\parallel \frac{m_-}{m_++m_-} \right)^2 \right)^{-2}.
\end{split}
\eeq

As a conclusion, the form of the coefficient of the wavefunction near two Dirac points can be written as
\beq
\label{wf_in-pl}
\begin{split}
|C_{(+k_{Fz}, \bk_\parallel), \bq}^{+-}|^2 &\approx \frac{8\pi^2\alpha^2 v_F E_B^0}{\Delta_Z^2 V} \left( 1+\frac{v_F^2q_z^2}{2(E_B^0)^2}  \right)^{-1} \left( E_B^0 + \frac{v_F^2q_z^2}{2E_B^0} + v_Fq_z + \frac{m_++m_-}{2m_+m_-} \left( \bk_\parallel + \bq_\parallel \frac{m_-}{m_++m_-} \right)^2 \right)^{-2}, \\
|C_{(-k_{Fz}, \bk_\parallel), \bq}^{+-}|^2 &\approx \frac{8\pi^2\alpha^2 v_F E_B^0}{\Delta_Z^2 V} \left( 1+\frac{v_F^2q_z^2}{2(E_B^0)^2}  \right)^{-1} \left( E_B^0 + \frac{v_F^2q_z^2}{2E_B^0} - v_Fq_z + \frac{m_++m_-}{2m_+m_-} \left( \bk_\parallel + \bq_\parallel \frac{m_-}{m_++m_-} \right)^2 \right)^{-2}, \\
\end{split}
\eeq
where $m_\pm^{-1} \equiv 2\alpha^2/\Delta_Z \pm m^{-1}$ and $E_B^0$ is given in Eq.~\er{ex}.

\section{Generation of spin-current by in-plane component of the tilted magnetic field for Dirac-point-vicinity model}
We now demonstrate that a finite spin-current can be generated by optically exciting the exciton-like collective mode in the presence of a tilted magnetic field. In this section, we will show that the in-plane component of the field is crucial for the generation of spin-current. 
The single-particle Hamiltonian can be written as
\beq
\label{H_d}
\hat{H} = \pm v_F k_\perp \hat{\sigma}_0 + \frac{k_\parallel^2}{2m} \hat{\sigma}_0 + \frac{\Delta_Z}{2} \hat{\sigma}_z + \frac{\Delta_y}{2} \hat{\sigma}_y + \alpha (k_y \hat{\sigma}_x - k_x \hat{\sigma}_y)
\eeq
where $\Delta_Z = g\mu_B B \cos\theta_\bB$ and $\Delta_y = g\mu_B B \sin\theta_\bB$, with $\theta_\bB$ as the polar angle of the magnetic field $\bB$. We assume large weight of the field to be along $\hat{z}$-direction, while a small weight along the plane, say $\hat{y}$, such that $\Delta_y \ll \Delta_Z$. 
The eigenvalues and eigenstates of the Hamiltonian \er{H_d} are given by
\beq
\label{es_tilt}
\begin{split}
\ve_\bk^s &= \pm v_F k_z + \frac{k_x^2 + k_y^2}{2m_b} + \frac{s}{2} \epsilon_{\bar{\bk}}, ~\epsilon_{\bar{\bk}} = \sqrt{4\alpha^2(\bar{k}_x^2 + k_y^2) + \Delta_Z^2}; \\
\ket{f_{\bk s}} &=
\left( \begin{array}{c}
\frac{i s \big( s \Delta_Z + \epsilon_{\bar{\bk}} \big)^{1/2}}{\sqrt{2\epsilon_{\bar{\bk}}}} \frac{(\bar{k}_x - ik_y)}{\sqrt{\bar{k}_x^2 + k_y^2}} \\
\frac{\big( -s \Delta_Z + \epsilon_{\bar{\bk}} \big)^{1/2}}{\sqrt{2 \epsilon_{\bar{\bk}}}}
\end{array} \right), \,\, \bar{k}_x = k_x - \frac{\Delta_y}{2\alpha},
\end{split}
\eeq
where $s=\pm1$ denotes the chirality of the spin-split subbands.
To discuss the spin-current, we first need to know how the dispersion of the collective mode and its wavefunction modify due to the presence of in-plane magnetic field. In order to do so, we repeat the same analysis as done in Secs.~\ref{sec:spin-ex} and \ref{sec:wf}, but now also in the presence of in-plane magnetic field.

\subsection{Dispersion of the collective mode frequency in the presence of tilted magnetic field}
\label{sec:dis_y}
Near the Dirac point, the matrix element of the integral equation \er{eq:full} for the exciton case ($s_1=+$ and $s_2=-$) decouples again and reduces to the same form as in Eq.~\er{bs}, except that $k_x$ only in the particle-hole excitation energy ($\epsilon_\bk$) is replaced by $\bar{k}_x$, where $\bar{k}_x$ is defined in Eq.~\er{es_tilt}. Accounting this change ($k_x \to \bar{k}_x$), we obtain the integral equation as
\beq
\label{ie_in-plane}
1 = -U \sum_{a=\pm} \int_{-\Delta_Z/2v_F}^{\Delta_Z/2v_F} \frac{dk_\perp}{2\pi} \int \frac{d^2k_\parallel}{(2\pi)^2} \frac{1}{\Omega_\bq - av_Fq_z - \frac{q_\parallel^2 + 2 \bk_\parallel \cdot \bq_\parallel}{2m} - \frac{1}{2} \epsilon_{\bar{\bk}_\parallel} - \frac{1}{2} \epsilon_{\bar{\bk}_\parallel+\bq_\parallel}},
\eeq
where
\beq
\label{ph_in-plane}
\epsilon_{\bar{\bk}_\parallel+\bq_\parallel} \approx \Delta_Z + \frac{4\alpha^2 \left( (\bar{k}_x+q_x)^2 + (k_y+q_y)^2 \right)}{2\Delta_Z}
\eeq
is the modified finite-$q$ particle-hole excitation energy in the presence of an in-plane magnetic field ($\Delta_y$), with $\bar{k}_\parallel = (\bar{k}_x^2 + k_y^2)^{1/2}$ and $\bar{k}_x$ given by Eq.~\er{es_tilt}. 
For simplicity, we do variable substitution, $k_x \to \bar{k}_x + \Delta_y/2\alpha$, and transform the integral in Eq.~\er{ie_in-plane} on $k_x$ to that on $\bar{k}_x$ ($k_y$ remains the same):
\beq
\label{dy}
\begin{split}
1 = -U \sum_{a=\pm} \int_{-\Delta_Z/2v_F}^{\Delta_Z/2v_F} \frac{dk_\perp}{2\pi} \int \frac{d^2\bar{k}_\parallel}{(2\pi)^2} \frac{1}{\Omega_\bq - \Delta_Z - av_Fq_z - \frac{m_++m_-}{2m_+m_-} \left( \bar{\bk}_\parallel + \bq_\parallel \frac{m_-}{m_++m_-} \right)^2 - \frac{q_\parallel^2}{2(m_++m_-)} - \frac{\Delta_y q_x}{2m\alpha}} \\
\end{split}
\eeq
As we can see, the in-plane magnetic field brings an extra last term in the denominator of the above integrand, as compared to that in the absence of it \er{bs}. 
We calculate the above integral \er{dy} in the same way as done in Eq.~\er{spin-ex}, and then solve for $\Omega_\bq$:
\beq
\label{dis_tilt}
\Omega_\bq \approx \Delta_Z - \frac{2\alpha^2 \Lambda^2}{\Delta_Z} \text{Exp} \left[ -\frac{8\pi^2 \alpha^2 v_F}{U \Delta_Z^2} \right] - \frac{v_F^2 q_z^2}{\frac{4\alpha^2 \Lambda^2}{\Delta_Z} \text{Exp} \left[ -\frac{8\pi^2 \alpha^2 v_F}{U \Delta_Z^2} \right]} + \frac{\Delta_y}{2m\alpha} q_x + \frac{(q_x^2 + q_y^2)}{2m} \frac{(4m^2\alpha^4 - \Delta_Z^2)}{4m\alpha^2 \Delta_Z}.
\eeq
The in-plane magnetic field leads to a linear in $q_x$ term in the dispersion, which tells us that at $q=0$ the exciton will move with velocity $\Delta_y/2m\alpha$ in the $\hat{x}$-direction. This is useful in the generation of spin-current which is the subject of discussion in Sec.~\ref{sec:spin-current}.

\subsection{Wavefunction of the collective mode frequency in the presence of tilted magnetic field}
We now discuss the wavefunction. As we saw, the dispersion along $\hat{y}$ and $\hat{z}$ remains unaffected by $\Delta_y$ \er{dis_tilt}. Hence, the $q_y$- and $q_z$-dependent parts of the wavefunction also will be unaffected by $\Delta_y$ and takes the same form as in Eq.~\er{wf_in-pl}. 
However, the $k_\parallel$ everywhere is replaced by $\bar{k}_\parallel$, which is where the dependence of $\Delta_y$ shows up in the wavefunction. 
Indeed, the explicit calculation, along the same lines as discussed in Secs.~\ref{sec:wf} and \ref{sec:dis_y}, suggests
\beq
\label{wf_in-pl}
\begin{split}
|C_{(+k_{Fz}, \bk_\parallel), \bq}^{+-}|^2 &\approx \frac{8\pi^2\alpha^2 v_F E_B^0}{\Delta_Z^2 V} \left( 1+\frac{v_F^2q_z^2}{2(E_B^0)^2}  \right)^{-1} \left( E_B^0 + \frac{v_F^2q_z^2}{2E_B^0} + v_Fq_z + \frac{m_++m_-}{2m_+m_-} \left( \bar{\bk}_\parallel + \bq_\parallel \frac{m_-}{m_++m_-} \right)^2 \right)^{-2}, \\
|C_{(-k_{Fz}, \bk_\parallel), \bq}^{+-}|^2 &\approx \frac{8\pi^2\alpha^2 v_F E_B^0}{\Delta_Z^2 V} \left( 1+\frac{v_F^2q_z^2}{2(E_B^0)^2}  \right)^{-1} \left( E_B^0 + \frac{v_F^2q_z^2}{2E_B^0} - v_Fq_z + \frac{m_++m_-}{2m_+m_-} \left( \bar{\bk}_\parallel + \bq_\parallel \frac{m_-}{m_++m_-} \right)^2 \right)^{-2}, \\
\end{split}
\eeq
where $|\bar{\bk}_\parallel| \equiv \bar{k}_\parallel = (\bar{k}_x^2 + k_y^2)^{1/2}$, with $\bar{k}_x \equiv k_x - \Delta_y/2\alpha$ given in Eq.~\er{es_tilt}, $m_\pm^{-1} \equiv 2\alpha^2/\Delta_Z \pm m^{-1}$ and $E_B^0$ is given in Eq.~\er{ex}. 
It can be checked that the wavefunction obeys the normalization condition: $\sum_\bk |C_{\bk\bq}^{+-}|^2 = 1$, or more explicitly $\sum_{\bk, a=\pm} |C_{(ak_{Fz},\bk_\parallel), \bq}^{+-}|^2 = 1$.

\subsection{Generation of spin-current}
\label{sec:spin-current}
In this section, we will discuss the generation of spin-current by in-plane magnetic field. More specifically, we will first calculate spin-expectation value in the exciton basis in Sec.~\ref{spin-exp} and find that the exciton carries a spin almost 1. 
Next, we will demonstrate the wavepacket representation of the exciton wavefunction in Sec.~\ref{wavepacket}. Finally, we will discuss the optical excitation of spin-excitons in Sec.~\ref{optical} which is needed to provide an estimate of the exciton generation rate. We will conclude this section by discussing that these excitons carry pure spin-current and any accompanying charge current is zero (Sec.~\ref{charge}).

\subsubsection{Derivation of spin-expectation value}
\label{spin-exp}
In the second quantized notation, the total spin operator is given by
\beq
\hat{\textbf{S}} = \sum_{\bk \alpha \alpha'} a_{\bk \alpha}^\dagger \textbf{S}_{\alpha\alpha'} a_{\bk\alpha'}, \,\,\,\, \text{where} \,\,\, \textbf{S}_{\alpha\alpha'} = \frac{1}{2} \langle f_{\bk\alpha}| \hat{\bs} |f_{\bk\alpha'} \rangle,
\eeq
where $| f_{\bk\alpha} \rangle$ is the single-particle eigenstate given by Eq.~\er{es} and $\hat{\bs}$ is a Pauli spin matrices. 
Using the form of spin-exciton state as 
\beq
\label{ex_q}
|xc(\bq)\rangle \equiv C_{\bk\bq}^{+-} a_{\bk+\bq,+}^\dagger a_{\bk-} |FG\rangle,
\eeq 
the expectation value of the spin operator can be written as
\beq
\label{sp}
\begin{split}
\langle xc(\bq)|\hat{\textbf{S}}|xc(\bq) \rangle = \frac{1}{2} \sum_{\bk_1 \bk_2 \bk_3} \sum_{\gamma_3\gamma_4} \langle FG| \left( C_{\bk_1 \bq}^{+-} \right)^* a_{\bk_1-}^\dagger a_{\bk_1+\bq,+} & a_{\bk_2 \gamma_3}^\dagger \langle f_{\bk_2 \gamma_3} |\hat{\bs}| f_{\bk_2 \gamma_4} \rangle a_{\bk_2 \gamma_4} \left( C_{\bk_3 \bq}^{+-} \right) a_{\bk_3+\bq,+}^\dagger a_{\bk_3-} |FG\rangle,
\end{split}
\eeq
As we can see, there are total six fermion operators. 
We now adopt operator RPA which amounts to using fermion anticommutation relation and collect only terms which are reduced to two fermion operators \cite{efimkin:2012}. 
Then we use the property: $\langle 0| a_{\bp \alpha}^\dagger a_{\bp \alpha'} |0 \rangle = n_{\bp \alpha} \delta_{\alpha \alpha'}$, where $n_{\bk \alpha}$ is the equilibrium Fermi function. 
Within RPA, we finally get \cite{efimkin:2012}
\beq
\label{spin-avg}
\langle xc(\bq)|\hat{\textbf{S}}|xc(\bq) \rangle = \frac{1}{2} \sum_\bk |C_{\bk_1 \bq}^{+-}|^2 \Big( \langle f_{\bk_1+\bq,+}|\hat{\bs}|f_{\bk_1+\bq,+} \rangle - \langle f_{\bk_1-}|\hat{\bs}|f_{\bk_1-} \rangle \Big) \left( n_{\bk_1-} - n_{\bk_1+\bq,+} \right).
\eeq
To derive Eq.~\er{spin-avg} we select creation-annihilation pairs in Eq.~\er{sp} and using the fermion anticommutation relation contract the number of fermion operators.
We start by selecting the pair $a_{\bk_2 \gamma_4} a_{\bk_3+\bq,+}^\dagger$ and write this as $\left( \delta_{\bk_2, \bk_3+\bq} \delta_{\gamma_4,+} - a_{\bk_3+\bq,+}^\dagger a_{\bk_2 \gamma_4} \right)$.
In the reduced four fermion operators, we select another such pair to reduce the same to two fermions. 
The averaging of two fermions then gives the first term of Eq.~\er{spin-avg}.
In the process of fermion reduction using its anticommutation relation, we will have two more terms with four fermions and six fermions.
The second term of Eq.~\er{spin-avg} comes from reducing the six fermions operator term to two fermions operator in the same way as it is done for the first term. 
So, finally we arrive at Eq.~\er{spin-avg} after averaging over the ground state Fermi gas $|FG\rangle$.
In the contracting process, we have multiple other terms with four fermion and six fermion operators. All these terms vanish upon averaging over the ground state.

Equation~\er{spin-avg} is the main equation as all of our results for the spin-expectation value follows from this equation only.
Using the forms of $|C_{\bk\bq}^{+-}|^2$ \er{wf_in-pl} and $|f_{\bp\pm}\rangle$ \er{es_tilt}, the spin-expectation value in the exciton basis \er{spin-avg} up to a leading order correction in $q$ can be calculated:
\beq
\begin{split}
\langle \hat{\textbf{S}}_z \rangle &= \frac{1}{2} \sum_\bk |C_{\bk\bq}^{+-}|^2 \left( \langle f_{\bk+\bq, +}|\hat{\sigma}_z|f_{\bk+\bq, +} \rangle - \langle f_{\bk -}|\hat{\sigma}_z|f_{\bk -} \rangle \right) \\
&= \frac{1}{2} \sum_{\bar{\bk}} |C_{\bar{\bk}\bq}^{+-}|^2 \left( \frac{\Delta_Z}{\epsilon_{\bar{\bk}+\bq_\parallel}} + \frac{\Delta_Z}{\epsilon_{\bar{\bk}}} \right) \approx \sum_{\bar{\bk}} |C_{\bar{\bk}\bq}^{+-}|^2 \left( 1 - \frac{\alpha^2}{\Delta_Z^2} \left( (\bar{\bk}_\parallel+\bq_\parallel)^2 + \bar{\bk}_\parallel^2 \right) \right) \\
&\approx 1 - \frac{\alpha^2(q_x^2+q_y^2)}{2\Delta_Z^2} \left( 1+\frac{\Delta_Z^2}{4m^2\alpha^4} \right) + \frac{v_F^2 q_z^2}{2\Delta_Z E_B^0} \text{log} \frac{2\alpha^2\Lambda^2}{\Delta_Z E_B^0} + \mathcal{O}(q^4), \\
\langle \hat{\textbf{S}}_x \rangle &= \frac{1}{2} \sum_\bk |C_{\bk\bq}^{+-}|^2 \left( \langle f_{\bk+\bq, +}|\hat{\sigma}_x|f_{\bk+\bq, +} \rangle - \langle f_{\bk -}|\hat{\sigma}_x|f_{\bk -} \rangle \right) \\
&= \frac{1}{2} \sum_{\bar{\bk}} |C_{\bar{\bk}\bq}^{+-}|^2 \left( \frac{2\alpha (k_y+q_y)}{\epsilon_{\bar{\bk}+\bq_\parallel}} + \frac{2\alpha k_y}{\epsilon_{\bar{\bk}}} \right) \\ 
&\approx -\frac{q_y}{2m\alpha} 
+ \mathcal{O}(q_\parallel^3), \\
\langle \hat{\textbf{S}}_y \rangle &= \frac{1}{2} \sum_\bk |C_{\bk\bq}^{+-}|^2 \left( \langle f_{\bk+\bq, +}|\hat{\sigma}_y|f_{\bk+\bq, +} \rangle - \langle f_{\bk -}|\hat{\sigma}_y|f_{\bk -} \rangle \right) \\
&= -\frac{1}{2} \sum_{\bar{\bk}} |C_{\bar{\bk}\bq}^{+-}|^2 \left( \frac{2\alpha (\bar{k}_x+q_x)}{\epsilon_{\bar{\bk}+\bq_\parallel}} + \frac{2\alpha \bar{k}_x}{\epsilon_{\bar{\bk}}} \right) \\ 
&\approx \frac{q_x}{2m\alpha} 
+ \mathcal{O}(q_\parallel^3).
\end{split}
\eeq 
As we can see, at $q=0$, $\langle \hat{\textbf{S}}_z \rangle \approx 1$, while $\langle \hat{\textbf{S}}_{x, y} \rangle$ vanishes. 
There is no $q_z$-dependent correction in $\langle \hat{S}_x \rangle$ and $\langle \hat{S}_y \rangle$ because the $\bk$-integral vanishes when $\bq \parallel \hat{z}$. Indeed, since $\epsilon_{\bar{\bk}+\bq_\parallel}$ does not depend on $q_z$, a simple observation suggests that the $\bk$-integral is odd.

\subsubsection{Wavepacket representation}
\label{wavepacket}
It is required to calculate $\langle xc(\bq)|\hat{\textbf{S}}(\br)|xc(\bq')\rangle$. Using Eq.~\er{ex_q} for the exciton wavefunction, and the extension of Eq.~\er{sp}, we can write
\beq
\begin{split}
\langle xc(\bq)|\hat{\textbf{S}}(\br)|xc(\bq')\rangle = \frac{1}{2} \sum_{\bk_1\bk_3} \sum_{\bk_2 \bk_2'} \sum_{\gamma_3\gamma_4} e^{-i(\bk_2 - \bk_2') \cdot \br} \langle FG| \left( C_{\bk_1 \bq}^{+-} \right)^* a_{\bk_1-}^\dagger a_{\bk_1+ \bq,+} a_{\bk_2 \gamma_3}^\dagger & \langle f_{\bk_2 \gamma_3} |\hat{\bs}| f_{\bk_2' \gamma_4} \rangle a_{\bk_2' \gamma_4} \\ 
\times & \left( C_{\bk_3 \bq'}^{+-} \right) a_{\bk_3+\bq',+}^\dagger a_{\bk_3-} |FG\rangle.
\end{split}
\eeq
Doing operator RPA the same way as in Eq.~\er{sp} to obtain Eq.~\er{spin-avg}, we obtain
\beq
\begin{split}
\langle xc(\bq)|\hat{\textbf{S}}(\br)|xc(\bq') \rangle = \frac{1}{2} e^{-i(\bq-\bq')\cdot \br} \sum_{\bk_1} \left( C_{\bk_1 \bq}^{+-} \right)^* \bigg( & \langle f_{\bk_1+\bq,+}|\hat{\bs}|f_{\bk_1+\bq',+} \rangle \left(C_{\bk_1 \bq'}^{+-}\right) \\
&- \langle f_{\bk_1+\bq-\bq',-}|\hat{\bs}|f_{\bk_1-} \rangle \left(C_{\bk_1+\bq-\bq', \bq'}^{+-} \right) \bigg) \big( n_{\bk_1-} - n_{\bk_1+\bq,+} \big),
\end{split}
\eeq
which reduces to 
\beq
\langle xc(\bq)|\hat{\textbf{S}}(\br)|xc(\bq') \rangle = \frac{1}{2} e^{-i(\bq-\bq')\cdot \br} \sum_{\bk_1} \left( C_{\bk_1 \bq}^{+-} \right)^* \bigg( \langle f_{\bk_1+\bq, +}|\hat{\bs}|f_{\bk_1+\bq', +} \rangle \left(C_{\bk_1 \bq'}^{+-}\right) - \langle f_{\bk_1+\bq-\bq', -}|\hat{\bs}|f_{\bk_1 -} \rangle \left(C_{\bk_1+\bq-\bq', \bq'}^{+-} \right) \bigg),
\eeq
near the Dirac point where we considered empty conduction ($n_{\bk+}=0$) and filled valence ($n_{\bk-}=1$) bands. 
Assuming small $q$ and $q'$, we get Eq.~(11) of the MT.

\subsubsection{Optical excitation of spin-excitons}
\label{optical}
In this section, we will provide technical details of the optical excitation of spin-excitons. 
The excitons form inside the spin gap near Dirac points of the Fermi surface. To excite them, one shines the light with frequency in resonance with the optical gap. According to the Fermi's golden rule, the transition probability in the long wavelength regime ($q\to 0$) of this excitation is given by $|\langle xc| \hat{\textbf{j}}\cdot\bA| FG \rangle|^2$, where $\hat{\textbf{j}}$ is current operator proportional to the velocity operator ($\hat{\bv}$)
\beq
\label{spin-cur}
\begin{split} 
\hat{\bv} = \left( \frac{k_x}{m} - \alpha \hat{\sigma}_y, \frac{k_y}{m} + \alpha \hat{\sigma}_x, \pm v_F \right),
\end{split}
\eeq
and $\bA$ is the vector potential. Since the main purpose of this step is to know the number of excitons excited per unit time, we assume for simplicity that $\Delta_y=0$. We calculate
\beq
\label{mat_el}
\begin{split}
\langle xc| \hat{\textbf{j}}\cdot\bA| FG \rangle &= \sum_{\bk\bk'} \sum_{\alpha\alpha'} \left(C_\bk^{+-}\right)^* \langle FG| a_{\bk-}^\dagger a_{\bk+} a_{\bk'\alpha}^\dagger\langle f_{\bk'\alpha}|\hat{\textbf{j}}\cdot\bA|f_{\bk'\alpha'}\rangle a_{\bk'\alpha'} |FG\rangle \\
&= \sum_{\bk\alpha} \left(C_\bk^{+-}\right)^* \langle FG| \left( \langle f_{\bk+}|\hat{\textbf{j}}\cdot\bA|f_{\bk\alpha}\rangle a_{\bk-}^\dagger a_{\bk\alpha} - \langle f_{\bk\alpha}|\hat{\textbf{j}}\cdot\bA|f_{\bk-}\rangle a_{\bk\alpha}^\dagger a_{\bk+} \right) |FG\rangle \\
&= \sum_\bk \left(C_\bk^{+-}\right)^* \langle f_{\bk+}|\hat{\textbf{j}}\cdot\bA|f_{\bk-}\rangle \left( n_{\bk-} - n_{\bk+} \right) \\
&= eU\alpha N^* \sum_\bk \left\{ \frac{k_-(\Delta_Z k_x + i\epsilon_\bk k_y)}{\epsilon_\bk(\Omega_0 - \epsilon_\bk)(k_x^2+k_y^2)} A_x + \frac{k_-(\Delta_Z k_y - i\epsilon_\bk k_x)}{\epsilon_\bk(\Omega_0 - \epsilon_\bk)(k_x^2+k_y^2)} A_y \right\} \left( n_{\bk-} - n_{\bk+} \right) \\
&= \frac{eU\alpha N^* \Delta_Z V}{4\pi^2 v_F} (A_x-iA_y) \int_0^\infty dk k \frac{\Delta_Z + \epsilon_\bk}{\epsilon_\bk (\Omega_0-\epsilon_\bk)} \\
&\approx \frac{eU\alpha N^* \Delta_Z V}{4\pi^2 v_F} (A_x-iA_y) \int_0^\infty dk k \frac{2\Delta_Z}{\Delta_Z (-E_B^0-2\alpha^2k^2/\Delta_Z)} \\
&\approx -\frac{eUN^*\Delta_Z^2 V}{8\pi^2\alpha v_F} (A_x-iA_y)\text{log}\frac{2\alpha^2\Lambda^2}{E_B^0\Delta_Z}, \,\,\,\, \text{where} \,\, \Lambda \gg \sqrt{\frac{E_B^0\Delta_Z}{2\alpha^2}}
\end{split}
\eeq
where $V$ is volume and we used Eq.~\er{wf} to write $C_\bk^{+-}$, with $N^* \equiv N_\bq^*(q=0)$ \er{norm} as the normalization factor. Also, in the above equation $\Lambda$ is the high-momentum cut off, $\Omega_0 \approx \Delta_Z-E_B^0$ (with $E_B^0\ll\Delta_Z$) is the exciton frequency at $q=0$ and $\epsilon_\bk$ is the particle-hole excitation energy given by Eq.~\er{ph_in-plane} at $q=0$ and $\Delta_y=0$ near Dirac points ($\pm k_F \hat{z}$) of the Fermi surface.
In the second last line of Eq.~\er{mat_el}, at leading order, the integrand is simplified as $(\Delta_Z+\epsilon_\bk)/\epsilon_\bk(\Omega_0-\epsilon_\bk) \approx -2/(E_B^0 + 2\alpha^2 k^2/\Delta_Z)$ according to the approximate form of $\epsilon_\bk$ \er{ph_in-plane}.
Finally, evaluating $|\langle xc| \hat{\textbf{j}}\cdot\bA| FG \rangle|^2$, with $|N|^2 = 8\pi^2\alpha^2 v_F E_B^0/V U^2\Delta_Z^2$ \er{N_0}, we get
\beq
\begin{split}
\Gamma = \frac{2\pi}{\hbar} |\langle xc| \hat{\textbf{j}}\cdot\bA| FG \rangle|^2 \delta(\Omega-\Omega_0) = \frac{e^2 \Delta_Z^2 E_B^0 V}{4\pi \hbar v_F} \left(\text{log}\frac{2\alpha^2\Lambda^2}{E_B^0\Delta_Z}\right)^2 \frac{E_x^2 + E_y^2}{\Omega^2} \delta(\Omega - \Omega_0),
\end{split}
\eeq
where $E_{x,y} = i\Omega A_{x,y}$ is the electric field component of electromagnetic (EM) wave.
To estimate the number of excitons per unit volume, we approximate delta-function by Lorentzian assuming finite lifetime ($\Delta^{-1}$) of excitons: $2\pi\delta(\Omega-\Omega_0) \approx [2\Delta/((\Omega-\Omega_0)^2 + \Delta^2)]_{\Delta\to 0}$.

We need to know the form of power ($P$) of the EM wave in terms of electric field. The energy flux per unit area per unit time is given by $\bS = (\bE \times \bB)/\mu_0$, where $\mu_0$ is the permeability of free space. The direction of $\bS$ is given by the direction of propagation of the EM wave. In free space, $B=E/c$, where $c$ is speed of light, we have $S=E^2/c\mu_0 = c\ve_0 E^2$, where we used $\ve_0\mu_0=1/c^2$. Upon time averaging, we get $\langle S \rangle = c\ve_0E^2/2$. Since $S$ is power per unit area, we can write $P = \bS \cdot \bA$, where $A$ is area. The average power is, therefore, $\langle P \rangle = c\ve_0 AE_\text{max}^2/2$, or $E_\text{max}^2 = 2\langle P \rangle/c\ve_0 A$. Upon recovering correct factors of $\hbar$, the optical excitation rate per unit volume ($\Gamma/V$) near the resonance ($\Omega \approx \Omega_0 \approx \Delta_Z$) can be written as 
\beq
\frac{\Gamma}{V} = \frac{2e^2E_B^0}{4\pi^2 \ve_0 c \hbar^2 v_F\Delta}\frac{P}{A} \left(\text{log}\frac{2\alpha^2\Lambda^2}{E_B^0\Delta_Z}\right)^2,
\eeq
where $E_B^0$ is the binding energy at $q=0$ given by Eq.~\er{ex} and $P/A$ is power per unit area. Here $\Lambda$ is the high-momentum cutoff introduced earlier in the definition of binding energy $E_B^0$ \er{ex}.

\subsubsection{Charge current}
\label{charge}
The charge current is defined as
\beq
\label{charge-cur}
I_c^{v_i} \propto \langle \hat{v}_j \rangle = \langle xc (\bq) | \hat{v}_i | xc (\bq) \rangle = \sum_\bk |C_{\bk\bq}^{+-}|^2 \left( \langle f_{\bk+\bq, +}|\hat{v}_i|f_{\bk+\bq, +} \rangle - \langle f_{\bk -}|\hat{v}_i|f_{\bk -} \rangle \right),
\eeq
where $\hat{\bv}$ is the velocity operator \er{spin-cur}.
As we know, the state $|f_{\bk\alpha} \rangle$ is a normalized eigenstate: $\langle f_{\bk\alpha}|f_{\bk\alpha} \rangle = 1$. Moreover, near Dirac points since the velocity along $\hat{z}$ direction is constant, $\hat{v}_z = \pm v_F$, the average of $\hat{v}_z$ for two bands cancels each other, and there is no charge current along $\hat{z}$. So, $I_c^{v_z} = 0$. 

Let us now discuss charge current in $\hat{x}$ and $\hat{y}$ directions.
Using Eqs.~\er{es_tilt}, \er{wf_in-pl} and \er{spin-cur}, we can write the expression for charge current along $\hat{x}$ direction as
\beq
\begin{split}
\label{icx}
I_c^{v_x} &\propto \sum_\bk |C_{\bk\bq}^{+-}|^2 \left[ \langle f_{\bk+\bq, +}| \left( \frac{k_x+q_x}{m} \hat{\sigma}_0 - \alpha \hat{\sigma}_y \right) |f_{\bk+\bq, +} \rangle - \langle f_{\bk -}| \left( \frac{k_x}{m} \hat{\sigma}_0 - \alpha \hat{\sigma}_y \right) |f_{\bk -} \rangle \right] \\
&= \frac{q_x}{m} + \alpha \sum_\bk |C_{\bk\bq}^{+-}|^2 \Big[ - \langle f_{\bk+\bq, +}| \hat{\sigma}_y |f_{\bk+\bq, +} \rangle + \langle f_{\bk -}| \hat{\sigma}_y |f_{\bk -} \rangle \Big] \,\,\,\, (\text{using normalization condition $\sum_\bk|C_{\bk\bq}^{+-}|^2=1$}) \\
&= \frac{q_x}{m} + \alpha \sum_\bk |C_{\bk\bq}^{+-}|^2 \left[ \frac{2\alpha(\bar{k}_x+q_x)}{\epsilon_{\bar{\bk}+\bq}} + \frac{2\alpha\bar{k}_x}{\epsilon_{\bar{\bk}}} \right] \\
&\approx \frac{q_x}{m} + \frac{2\alpha^2}{\Delta_Z} \sum_\bk |C_{\bk\bq}^{+-}|^2 \left( 2\bar{k}_x + q_x \right) \,\,\,\, (\text{using $\epsilon_\bk \approx \Delta_Z$ \er{ph_in-plane} at leading order}) \\
&= \frac{q_x}{m_+} + \frac{8\pi^2\alpha^2 v_F E_B^0}{\Delta_Z^2} \left(1+\frac{v_F^2q_z^2}{2(E_B^0)^2}\right)^{-1} \frac{4\alpha^2}{\Delta_Z} \sum_{\bk, a=\pm} \frac{\bar{k}_x}{\left\{ E_B^0 + \frac{v_F^2q_z^2}{2E_B^0} + av_Fq_z + \frac{2m_+m_-}{m_++m_-} \left( \bar{\bk}_\parallel+\bq_\parallel\frac{m_-}{m_++m_-}\right)^2 \right\}^2} \\
&\text{In the second term, the small-$q$ expansion leads to the finite contribution only from $\bar{k}_x q_x$ term} \\
&\Rigt I_c^{v_x} \propto \frac{q_x}{m_+} - \frac{q_x}{m_+} \left(1+\frac{v_F^2q_z^2}{2(E_B^0)^2}\right)^{-1} = 0.
\end{split}
\eeq
In the last line of the above equation, the correction term is an artifact of small-$q$ expansion which results into beyond-quadratic-dispersion approximation term. Hence it is safely ignored.
The calculation along same lines for the charge current along $\hat{y}$-direction, it can be verified that $I_c^{v_y}=0$ as well.

\subsection{Derivation of X-wave solution of the hyperbolic Schrodinger equation}
In this section, we will provide technical details of the derivation of envelope $X$ wave solution (see Eq.~(13) of the MT) of the hyperbolic Schrodinger equation (HSE) given by Eq.~(11) of the MT. 
In Eq.~(11) of the MT, we make the transformation, $x, y, z \to x'\sqrt{p_1}, y'\sqrt{p_1}, z'\sqrt{p_2}$, and write the HSE as
\beq
\label{SE1}
\left( i\frac{\partial}{\partial t} - \Omega_0 \right) \Psi(\br', t) = \left[ -\left( \frac{\partial^2}{\partial x'^2} + \frac{\partial^2}{\partial y'^2} \right) + \frac{\partial^2}{\partial z'^2} - \frac{ip_3}{\sqrt{p_1}} \frac{\partial}{\partial x'} \right] \Psi(\br', t).
\eeq
We use the following ansatz in the above equation: $\Psi(\br', t) = e^{-i\Omega_0 t} e^{-i p_3x'/2\sqrt{p_1}} e^{ip_3^2t/4p_1} \phi(x', y', z', t)$. The HSE \er{SE1} reduces to
\beq
i\frac{\partial \phi}{\partial t} = -\left( \frac{\partial^2}{\partial x'^2} + \frac{\partial^2}{\partial y'^2} \right) \phi + \frac{\partial^2 \phi}{\partial z'^2}.
\eeq
To solve this differential equation, we use $\phi(x', y', z', t) = \varphi(\xi', y', z') e^{iv'x'/2} e^{-itv'^{2}/4}$, where $\xi' \equiv x'-v't$, with $v' \equiv v/\sqrt{p_1}$ as the wave propagating velocity, and obtain
\beq
\label{de1}
\left[\frac{\partial^2}{\partial \xi'^2} + \frac{\partial^2}{\partial y'^2} - \frac{\partial^2}{\partial z'^2} \right] \varphi (\xi', y', z') = 0.
\eeq
The non-spreading solution of this equation has been discussed in detail in Ref.~\cite{stefano:2004}, but with the propagating velocity ($v$) assumed to be along $\hat{z}$. Extending the analysis to our case, we can write the general solution of Eq.~\er{de1}:
\beq
\varphi(\xi', y', z') = \int_0^\infty d\alpha f(\alpha) J_0(\alpha r_\perp') e^{i\alpha z'}, \,\,\, \text{where} \,\,\, r_\perp' \equiv (\xi'^2 + y^2)^{1/2}.
\eeq
Finally, the full solution of Eq.~\er{SE1} can be written as
\beq
\label{x-wave}
\Psi(\br', t) = e^{-i\Omega_0t} e^{-ip_3x'/2\sqrt{p_1}} e^{ip_3^2t/4p_1} e^{iv'x'/2} e^{-itv'^2/4} \int_0^\infty d\alpha f(\alpha) J_0(\alpha r_\perp') e^{i\alpha z'},
\eeq
which, in the original cartesian basis $(x, y, z, t)$ can be written as
\beq
\label{x-wave_org}
\begin{split}
\Psi(\br, t) = e^{-ip_3 (x-p_3t/2)/2p_1} e^{iv(x-vt/2)/2p_1} e^{-i\Omega_0t} \int_0^\infty d\alpha f(\alpha) J_0(\alpha r_\parallel/\sqrt{p_1}) e^{i\alpha z/\sqrt{p_2}},
\end{split}
\eeq
where $r_\parallel \equiv \big( (x-vt)^2 + y^2 \big)^{1/2}$. 
As we can see, in the above solution \er{x-wave_org} the velocity $v$ is completely arbitrary. 
For an exponentially decaying spectrum $f(\alpha) = e^{-\alpha\gamma}$ \cite{stefano:2004, huang:2020}, where $\gamma$ is some arbitrary width, if we assume $v = p_3 \equiv \Delta_y/2m\alpha$, we obtain a fundamental linear $X$ wave solution:  
\beq
\Psi_X(\br, t) = \sqrt{p_1p_2}e^{-i\Omega_0 t}\left[ p_2(x-p_3t)^2 + p_2y^2 + p_1(\sqrt{p_2}\gamma - i z)^2 \right]^{-1/2},
\eeq
as given by Eq.~(11) of the MT. 
This is a shape preserving envelope $X$ wave solution of the HSE \er{SE1} which propagates with velocity $v_{xc}=p_3=\Delta_y/2m\alpha$ in the $\hat{x}$ direction.
We note that the above general $X$-wave solution \er{x-wave_org} is obtained for the linear HSE. The general $X$ wave solution for the non-linear HSE has been discussed in Refs.~\cite{stefano:2004, trull:2003, nikolaos:2009, sedov:2015, magdalena:2016, antonio:2018, huang:2020}, which is beyond the scope of our work.

\section{Spherical Fermi surface model}
\label{SFS}
In the diagrammatic RPA approach we consider tilted magnetic field such that its components along one of the planar directions, say $\hat{y}$, in addition to the $\hat{z}$-direction as considered in the previous section (Sec.~\ref{model}), is also finite.
The single-particle Hamiltonian for the system can be written as
\beq
\label{ham}
\hat{H} = \frac{k^2}{2m} \hat{\sigma}_0 + \frac{\Delta_Z}{2} \hat{\sigma}_z + \frac{\Delta_y}{2} \hat{\sigma}_y + \alpha (k_y \hat{\sigma}_x - k_x \hat{\sigma}_y),
\eeq
where $m$ is the band mass, $\sigma_i$ is the Pauli matrix in the spin space, $k$ is the electron momenta, and $\Delta_Z = g \mu_B B \cos\theta_B$ and $\Delta_y = g \mu_B B \sin\theta_B \sin\phi_B$, with $g$ as the electron Lande factor, $\mu_B$ as the Bohr magneton, $B$ as the magnitude of the magnetic field and $\theta_B (\phi_B)$ as the polar (azimuthal) angle of magnetic field, are Zeeman energy splittings due to magnetic field components along $z$ and $y$ directions, respectively. Note that the polar $(\theta_B)$ and azimuthal $(\phi_B)$ angles of magnetic field are different from those of electronic momenta which will be denoted as $\theta_\bk$ and $\phi_\bk$, respectively. To avoid further confusion, we will use $\Delta_Z$ and $\Delta_y$ for perpendicular and in-plane magnetic fields, correspondingly, from now onwards. We will treat the electron-electron interaction at short ranges \er{hub}.

The eigenvalues and eigenstates of the single-particle Hamiltonian \er{ham} are given by
\beq
\label{es_sfs}
\begin{split}
\ve_\bk^s &= \frac{k^2}{2m} + \frac{s}{2} \epsilon_{\bar{\bk}}, ~\epsilon_{\bar{\bk}} = \sqrt{4\alpha^2(\bar{k}_x^2 + k_y^2) + \Delta_Z^2}; \\
\ket{s} &=
\left( \begin{array}{c}
\frac{i s \big( s \Delta_Z + \epsilon_{\bar{\bk}} \big)^{1/2}}{\sqrt{2\epsilon_{\bar{\bk}}}} \frac{(\bar{k}_x - ik_y)}{\sqrt{\bar{k}_x^2 + k_y^2}} \\
\frac{\big( -s \Delta_Z + \epsilon_{\bar{\bk}} \big)^{1/2}}{\sqrt{2 \epsilon_{\bar{\bk}}}}
\end{array} \right), \bar{k}_x = k_x - \frac{\Delta_y}{2\alpha},
\end{split}
\eeq
where $s=\pm1$ denotes the chirality of the spin-split subbands. Notice that the eigenstate for the spherical Fermi surface (SFS) model \er{es_sfs} is same as that for the DPV model \er{es_tilt}.

The collective modes in the spin sector of an interacting system show up as poles of the full spin susceptibility defined as
\beq
\chi_{ij} (i\Omega_n, \bq) = -\Pi_{ij}^U(i\Omega_n, \bq),
\eeq
where
\beq
\label{full_bubble}
\Pi_{ij}^U(Q) = \int_K \text{Tr} \big[ \hat{\sigma}_i \hat{G}^0(K) \hat{\gamma}_j \hat{G}^0(K+Q) \big]
\eeq
is the full spin polarization bubble, with $Q = (i\Omega_n, \bq)$, $\int_K \equiv T\sum_{\omega_m} \int d^3k/(2\pi)^3$, $\epsilon_m$ as the Matsubara frequency, $\hat{\gamma}_i$ as the $2\times2$ spin-interaction vertex,
\beq
\label{gr}
\begin{split}
\hat{G}^0(i\epsilon_m, & \bk) = \sum_s \frac{\ket{s} \bra{s}}{i\epsilon_m + \mu - \ve_\bk^s} = \sum_s \hat{D}_s(\bk) g(i\epsilon_m, \bk), \\
\hat{D}_s(\bk) &= \frac{1}{2} \bigg[ \hat{\sigma}_0 + s \frac{\Delta_Z}{\epsilon_{\bar{\bk}}} \hat{\sigma}_z + s \frac{2\alpha}{\epsilon_{\bar{\bk}}} (\hat{\sigma}_x k_y - \hat{\sigma}_y \bar{k}_x) \bigg], \\
g_s(i\epsilon_m, & \bk) = \frac{1}{i\epsilon_m + \mu - \ve_\bk^s}
\end{split}
\eeq
as the single-particle Green's function and $\mu$ as the chemical potential. Here, we term $g_s(i\epsilon_m, \bk)$ as a chiral Green's function. The eigenvalues ($\ve_\bk^s$), eigenstates ($\ket{s}$) and $\epsilon_\bk$ appearing in the definition of Green's function \er{gr} are given in Eq.~\er{es_sfs}. To obtain the full spin susceptibility, we perform a random phase approximation (RPA) which sums up polarization bubbles containing ladder series of vertex corrections. For a doped 3D semiconductor, the interaction line inside the bubble is supposed to be a screened Coulomb potential. However, a closed form of the full polarization bubble \er{full_bubble} for a screened Coulomb potential cannot be obtained analytically; therefore, we adopt an approximation in which the interaction is replaced by a momentum-independent constant (short-range Hubbard $U$) \cite{maiti_2014}. Within this assumption, the spin-interaction vertex $\hat{\gamma}_i$ can be written as
\beq
\label{ver}
\hat{\gamma}_j = \sigma_j - U \int_K \hat{G}^0(K) \hat{\gamma}_j \hat{G}(K+Q).
\eeq
We expand $\hat{\gamma}_j$ over a completet set of Pauli matrices
\beq
\label{exp}
\hat{\gamma}_j = M_j^a \hat{\sigma}_a,
\eeq
where $a \in 0,1,2,3$ and the coefficient $M_j^a$ forms a $4\times4$ matrix. Substituting Eq.~\er{exp} back into Eq.~\er{ver}, one can find
\beq
\label{M}
\begin{split}
\hat{\Pi}^U &= \hat{\Pi}^0 \hat{M}, \\
\hat{M} &= \bigg( \hat{I} + \frac{U}{2} \hat{\Pi}^0 \bigg)^{-1},
\end{split}
\eeq
where $\hat{\Pi}_0$ is the non-interacting bubble and $\hat{I}$ is the $4\times4$ identity matrix. Collective modes are given by poles of $\Pi_{ij}^U$ which are solutions of $\text{Det}(\hat{M}^{-1}) = 0$.

\subsection{Chiral spin resonances}
We will first discuss $q=0$ spin collective modes for the case when $\Delta_y=0$. Later we will include small $\Delta_y$ such that $\Delta_y \ll \Delta_Z, \Delta_R \ll \mu$, where $\Delta_R = 2\alpha k_F$ is the Rashba energy. We will see later that there are two Rashba energy scales in the system, $\Delta_R$ and $m\alpha^2$ with $m\alpha^2 \ll \Delta_R$, and $\Delta_y$ will be assumed to be small compared to both the energy scales. The finite $q$ case will be discussed assuming $q^2/2m \ll v_F q \ll \Delta_y$, the main goal of which is to track the dispersion of a particular chiral spin mode which is robust and remains exponentially close to the lower edge of the spin-flip particle-hole continuum.

As mentioned in Eq.~\er{M}, to know collective modes, we require the form of all non-zero components of the non-interacting spin polarization bubble (assuming spherical FS and $q=0$ and $\Delta_y=0$):
\bwt
\beq
\label{Pi}
\begin{split}
\Pi_{xx} (i\Omega_n) &= -\frac{\nu_F}{4} \Bigg( 2 + \frac{(-\Omega_n^2 + \Delta_Z^2)}{\Delta_R \sqrt{\Omega_n^2 + \Delta_R^2 + \Delta_Z^2}} \text{Log} \Bigg[ \frac{\sqrt{\Omega_n^2 + \Delta_R^2 + \Delta_Z^2} + \Delta_R}{\sqrt{\Omega_n^2 + \Delta_R^2 + \Delta_Z^2} - \Delta_R} \Bigg] \Bigg), \\
\Pi_{yy} (i\Omega_n) &= \Pi_{xx} (i\Omega_n), \\
\Pi_{xy} (i\Omega_n) &= \frac{\nu_F}{4} \frac{2\Omega_n \Delta_Z}{\Delta_R \sqrt{\Omega_n^2 + \Delta_R^2 + \Delta_Z^2}} \text{Log} \Bigg[ \frac{\sqrt{\Omega_n^2 + \Delta_R^2 + \Delta_Z^2} + \Delta_R}{\sqrt{\Omega_n^2 + \Delta_R^2 + \Delta_Z^2} - \Delta_R} \Bigg], \\
\Pi_{yx} (i\Omega_n) &= -\Pi_{xy} (i\Omega_n), \\
\Pi_{zz} (i\Omega_n) &= -\frac{\nu_F}{4} \Bigg( 4 - \frac{2(\Omega_n^2 + \Delta_Z^2)}{\Delta_R \sqrt{\Omega_n^2 + \Delta_R^2 + \Delta_Z^2}} \text{Log} \Bigg[ \frac{\sqrt{\Omega_n^2 + \Delta_R^2 + \Delta_Z^2} + \Delta_R}{\sqrt{\Omega_n^2 + \Delta_R^2 + \Delta_Z^2} - \Delta_R} \Bigg] \Bigg),
\end{split}
\eeq
\ewt
where $\nu_F \equiv mk_F/\pi^2$ is the 3D density of states assuming spherical FS.
While obtaining above bubbles \er{Pi}, we assumed the spin splitting energy ($\epsilon_{\bar{\bk}}$) to be small compared to the Fermi energy such that the Fermi function, $n_F(\xi_\bk + s/2 \epsilon_{\bar{\bk}})$, which appears after frequency summation in the form of the bubble (actually a difference of Fermi functions, corresponding to transition between two spin-split subbands, shows up: $n_F(\xi_\bk + \epsilon_{\bar{\bk}}/2) - n_F(\xi_\bk - \epsilon_{\bar{\bk}}/2)$) can be expanded as $n_F(\xi_\bk + (s/2) \epsilon_{\bar{\bk}}) \approx n_F(\xi_\bk) + (s/2) \epsilon_{\bar{\bk}} n_F'(\xi_\bk)$, where $\xi_\bk = k^2/2m - \mu$. At $T=0$, $n_F(\xi_\bk) \approx \Theta(-\xi_\bk)$ and $n_F'(\xi_\bk) \approx -\delta(\xi_\bk)$. 
One can see that all the bubbles above \er{Pi} show resonances at $\Omega = \Delta_z$ and $\Omega = \sqrt{\Delta_z^2 + \Delta_R^2}$, where $\Omega$ is the retarded frequency, which are identified as lower and upper edges of the spin-flip single-particle continuum. The form of the coefficient matrix $\hat{M}$ suggests decoupling of the in-plane, \{x, y\}, and out-of-plane, \{z\}, sectors:
\beq
\label{det}
\text{Det}[\hat{M}^{-1}] = \bigg( 1 + \frac{U}{2} \Pi_{zz} \bigg) \bigg[ \bigg( 1 + \frac{U}{2} \Pi_{xx} \bigg)^2 + \frac{U^2}{4} \Pi_{xy}^2 \bigg] = 0.
\eeq
At weak coupling ($u\ll1$, where $u \equiv U\nu_F/4 > 0$), there exists only one mode, in the in-plane sector, which is robust and remains exponentially close to the lower edge of the continuum, as given by Eq.~(11) of the MT. 
The mode in the out-of-plane sector, on the other hand, is Landau damped at weak coupling; it is peeled off the continuum only at strong enough coupling which is beyond the regime of validity of our theory. 
The exact analytical form of collective modes from Eq.~\er{det} cannot be calculated because of the transcendental nature of the equation for $\Omega$ that one would encounter at some point. However, near the lower edge of the continuum, i.e., $\Omega \approx \Delta_Z$, the analytical form of this mode can certainly be obtained by assuming $\Omega = \Delta_Z - E_B$, with $E_B \ll \Delta_Z$ as the binding energy. 
Expanding Eq.~\er{Pi} for small $E_B$, the form of bubbles \er{Pi} (up to Log singularity) can be written as
\beq
\label{Pi_app}
\begin{split}
\Pi_{xx} &\approx \frac{\nu_F}{4} \bigg( -2 + \frac{2\Delta_Z^2}{\Delta_R^2} \text{Log} \frac{\Delta_Z E_B}{2 \Delta_R^2} \bigg), \\
\Pi_{xy} &\approx \frac{\nu_F}{4} \bigg( \frac{2i\Delta_Z^2}{\Delta_R^2} \text{Log} \frac{\Delta_Z E_B}{2 \Delta_R^2} \bigg), \\
\Pi_{zz} &\approx \frac{\nu_F}{4} \bigg( -4 - \frac{4\Delta_Z E_B }{\Delta_R^2} \text{Log} \frac{\Delta_Z E_B}{2 \Delta_R^2} \bigg).
\end{split}
\eeq
As one can see that for $E_B \to 0$, the Log in the above equation diverges. One can immediately find from the formula of $\Pi_{zz}$ in Eq.~\er{Pi_app} that the Log-singular term is vanishingly small due to the multiplicative factor $E_B$, unlike that in the planar components of the bubble. This tells us that for the mode to be well lived in this sector, the coupling $u$ has to be strong to reinforce the smallness imposed by the binding energy. But then at strong coupling, the in-plane mode tends to become soft indicating ferromagnetic instability (in our case of magnetized polar metals, this instability could be ferroelectric-like), as discussed in the context of 2D semiconductor heterostructures \cite{maiti2016, kumar2017}. So, we remain at weak coupling and consider only the mode that exists in the in-plane sector below the continuum; the out of plane mode exists inside the continuum at weak coupling and is Landau damped. In the in-plane sector, essentially we are required to solve
\beq
\label{plane}
1 + \frac{U}{2} \Pi_{xx} = \pm i \frac{U}{2} \Pi_{xy}
\eeq
according to Eq.~\er{det}. The one with the minus sign in Eq.~\er{plane} does not result in any mode because the term containing leading Log-singularity cancels identically. The one with the plus sign gives collective mode as written in Eq.~(11) of the MT. For further convenience, let's rewrite the form of this mode explicitly:
\beq
\label{mode}
\Omega \approx \Delta_Z - \frac{2\Delta_R^2}{\Delta_Z} \text{Exp} \Big[ -\frac{\Delta_R^2(1-u)}{\Delta_Z^2 2u} \Big].
\eeq
Note that one cannot recover Silin-Leggett mode by substituting $\Delta_R=0$ in the above frequency \er{mode}: while writing \er{Pi_app}, we assumed $E_B$ as the smallest energy scale such that in addition to $E_B\ll\Delta_Z$ we also have $E_B\ll\Delta_R$. Putting $\Delta_R=0$ in Eq.~\er{mode} will simply violate our assumption, $E_B\ll\Delta_Z, \Delta_R$.

\subsection{Chiral spin waves in the presence of tilted magnetic field}
We now calculate the dispersion of the collective mode \er{mode} at small $q$. 
Moreover, we now also turn on the in-plane component of the magnetic field, $\Delta_y$, assuming $\Delta_y \ll \Delta_Z, \Delta_R$. 
At finite $q$ and $\Delta_y$, all the components of $\hat{\Pi}^0$, including the one in the charge sector, are non-zero. 
Since now $\Delta_y$ is non-zero, the mirror symmetry along $\hat{x}$ is broken. The dispersion of the collective mode along $\hat{x}$- and $\hat{y}$-directions, therefore, will not be the same, same as what we saw for the DPV model \er{dis_tilt}. 
Since $\Delta_y$ gives a boost along $\hat{x}$, see the single-particle Hamiltonian \er{ham} and its eigenvalue \er{es_tilt} where $\Delta_y$ dresses up $k_x$ to $\bar{k}_x = k_x - \Delta_y/2\alpha$, the dispersion along $\hat{x}$ should now start off linearly. In the $\hat{y}$-direction, however, it should still start off quadratically only, as discussed in the context of DPV model (Sec.~\ref{sec:dis_y}). 
The linear dispersion along $\hat{x}$ can be understood from symmetry: the presence of $\Delta_y$ ensures broken mirror symmetry along the $\hat{x}$ which explains the origin of boost along $\hat{x}$-direction in the dispersion. 
Since the mirror symmetry along $\hat{y}$ is preserved, there is no boost and, therefore, the dispersion should be quadratic. In the following, we will calculate the dispersion of collective mode by first considering $\bq \parallel \hat{x}$ (Sec.~\ref{q_x}), followed by $\bq \parallel \hat{y}$ (Sec.~\ref{q_y}) and then $\bq \parallel \hat{z}$ (Sec.~\ref{q_z}).

\subsubsection{$\bq \parallel \hat{x}$}
\label{q_x}
The bare polarization bubble $\hat{\Pi}^0$ at arbitrary $q$, assuming it to be along $x$-direction, can be written as
\beq
\begin{split}
\Pi_{ij} (\bq, i\Omega_n) &= \frac{1}{2} \sum_{s, \bar{s}} \int_K f_{ij} g_{\bar{s}}(i\omega_m, \bk) g_s \big( i(\omega_m + \Omega_n), \bk+\bq \big), \\
&= \frac{1}{2} \sum_{s, \bar{s}} \int_K f_{ij} \frac{1}{i \omega_m - \xi_\bk^{\bar{s}}} \frac{1}{i(\omega_m + \Omega_n) - \xi_{\bk+\bq}^s},
\end{split}
\eeq
where $\xi_\bk^s = \ve_\bk^s - \mu$, with $\ve_\bk^s$ as the eigenenergy \er{es_sfs}, and $f_{ij}$ is the coherence factor calculated as
\bwt
\beq
\label{coh}
\begin{split}
f_{00} &= 1 + s\bar{s} \frac{\Delta_Z^2 + 4\alpha^2 \big( \bar{k}_x(\bar{k}_x+q) + k_y^2 \big)}{\epsilon_{\bar{k}_x, k_y} \epsilon_{\bar{k}_x+q, k_y}}, \\
f_{0x} &= s \frac{2\alpha k_y}{\epsilon_{\bar{k}_x+q, k_y}} + \bar{s} \frac{2\alpha k_y}{\epsilon_{\bar{k}_x, k_y}} - is \bar{s} \frac{\Delta_Z 2\alpha q}{\epsilon_{\bar{k}_x, k_y} \epsilon_{\bar{k}_x+q, k_y}}, \\
f_{x0} &= f_{0x}(i \rigt -i), \\
f_{0y} &= - s \frac{2\alpha (\bar{k}_x+q)}{\epsilon_{\bar{k}_x+q, k_y}} - \bar{s} \frac{2\alpha \bar{k}_x}{\epsilon_{\bar{k}_x, k_y}}, \\
f_{y0} &= f_{0y}, \\
f_{0z} &= (s+\bar{s}) \frac{\Delta_Z}{\epsilon_{\bar{k}_x+q, k_y}} + is \bar{s} \frac{4\alpha^2 k_y q}{\epsilon_{\bar{k}_x, k_y} \epsilon_{\bar{k}_x+q, k_y}}, \\
f_{z0} &= f_{0z}(i \rigt -i), \\
f_{xx} &= 1 - s\bar{s} \frac{\Delta_Z^2}{\epsilon_{\bar{k}_x, k_y} \epsilon_{\bar{k}_x+q, k_y}} + s\bar{s} \frac{4\alpha^2 \big( k_y^2 - \bar{k}_x(\bar{k}_x+q) \big)}{\epsilon_{\bar{k}_x, k_y} \epsilon_{\bar{k}_x+q, k_y}}, \\
f_{yy} &= 1 - s\bar{s} \frac{\Delta_Z^2}{\epsilon_{\bar{k}_x, k_y} \epsilon_{\bar{k}_x+q, k_y}} - s\bar{s} \frac{4\alpha^2 \big( k_y^2 - \bar{k}_x(\bar{k}_x+q) \big)}{\epsilon_{\bar{k}_x, k_y} \epsilon_{\bar{k}_x+q, k_y}}, \\
f_{zz} &= 1 + s\bar{s} \frac{\Delta_Z^2}{\epsilon_{\bar{k}_x, k_y} \epsilon_{\bar{k}_x+q, k_y}} - s\bar{s} \frac{4\alpha^2 \big( k_y^2 + \bar{k}_x(\bar{k}_x+q) \big)}{\epsilon_{\bar{k}_x, k_y} \epsilon_{\bar{k}_x+q, k_y}}, \\
f_{xy} &= is \frac{\Delta_Z}{\epsilon_{\bar{k}_x+q, k_y}} - i\bar{s} \frac{\Delta_Z}{\epsilon_{\bar{k}_x, k_y}} - s \bar{s} \frac{4\alpha^2 \big( k_y (\bar{k}_x + q) + \bar{k}_x k_y \big)}{\epsilon_{\bar{k}_x, k_y} \epsilon_{\bar{k}_x+q, k_y}}, \\
f_{yx} &= f_{xy} (i \rigt -i), \\
f_{xz} &= is \frac{2\alpha (\bar{k}_x + q)}{\epsilon_{\bar{k}_x+q, k_y}} - i\bar{s} \frac{2\alpha \bar{k}_x}{\epsilon_{\bar{k}_x, k_y}} + s \bar{s} \frac{2\Delta_Z 2\alpha k_y}{\epsilon_{\bar{k}_x, k_y} \epsilon_{\bar{k}_x+q, k_y}}, \\
f_{zx} &= f_{xz} (i \rigt -i), \\
f_{yz} &= is \frac{2\alpha k_y}{\epsilon_{\bar{k}_x+q, k_y}} - i\bar{s} \frac{2\alpha k_y}{\epsilon_{\bar{k}_x, k_y}} - s \bar{s} \frac{\Delta_Z 2\alpha (2\bar{k}_x+q)}{\epsilon_{\bar{k}_x, k_y} \epsilon_{\bar{k}_x+q, k_y}}, \\
f_{zy} &= f_{yz} (i \rigt -i).
\end{split}
\eeq
\ewt
In the regime $q^2/2m \ll v_F q \ll \Delta_y \ll \Delta_Z, \Delta_R$, the form of all the bubbles near the lower edge of the spin-flip continuum ($\Omega = \Delta_Z$ is still the location of the lower edge of the continuum even at finite $\Delta_y$ because of our assumption $\Delta_y \ll \Delta_R$ \cite{kumar2023}) up to leading singularity can be written as
\bwt
\beq
\begin{split}
\Pi_{00} &= \frac{\nu_F}{8} \Bigg[ q^2 \frac{4\alpha^2}{\Delta_R^2} \text{Log} \frac{\Delta_Z E_B}{2\Delta_R^2} + q^2 \Delta_y^2 \frac{4\alpha^2}{2\Delta_R^4} \text{Log} \frac{\Delta_Z E_B}{2\Delta_R^2} \Bigg], \\
\Pi_{0x} &= \frac{\nu_F}{8} \Bigg[ iq \frac{4\alpha \Delta_Z}{\Delta_R^2} \text{Log} \frac{\Delta_Z E_B}{2\Delta_R^2} + iq \Delta_y^2 \frac{4\alpha \Delta_Z}{2\Delta_R^4} \text{Log} \frac{\Delta_Z E_B}{2\Delta_R^2} \Bigg], \\
\Pi_{x0} &= -\Pi_{0x}, \\
\Pi_{0y} &= \frac{\nu_F}{8} \Bigg[ -q \frac{4\alpha \Delta_Z}{\Delta_R^2} \text{Log} \frac{\Delta_Z E_B}{2\Delta_R^2} - q \Delta_y^2 \frac{4\alpha \Delta_Z}{2\Delta_R^4} \text{Log} \frac{\Delta_Z E_B}{2\Delta_R^2} \Bigg], \\
\Pi_{y0} &= \Pi_{0y}, \\
\Pi_{0z} &= \frac{\nu_F}{8} \Bigg[ -iq \frac{4\alpha}{\Delta_R} \text{Log} \frac{\Delta_Z E_B}{2\Delta_R^2} - iq \Delta_y^2 \frac{4\alpha}{2\Delta_R^3} \text{Log} \frac{\Delta_Z E_B}{2\Delta_R^2} \Bigg], \\
\Pi_{z0} &= -\Pi_{0z}, \\
\Pi_{xx} &= \frac{\nu_F}{8} \Bigg[ \bigg( -4 + \frac{4\Delta_Z^2}{\Delta_R^2} \text{Log} \frac{\Delta_Z E_B}{2\Delta_R^2} \bigg) + \Delta_y^2 \bigg( \frac{2\Delta_Z^2}{\Delta_R^4} \text{Log} \frac{\Delta_Z E_B}{2\Delta_R^2} \bigg) \\
&\hspace{1cm} + \frac{v_F q \Delta_y}{E_B} \frac{4\Delta_Z^2}{\Delta_R^3} + \frac{2v_F^2 q^2}{\Delta_R^2} \frac{\Delta_Z (4m^2\alpha^4 - \Delta_Z^2)}{E_B \Delta_R^2} + \mathcal{O} (q^2 \Delta_y^2) \Bigg], \\
\Pi_{yy} &= \Pi_{xx}, \\
\Pi_{zz} &= \frac{\nu_F}{8} \Bigg[ \bigg( -8 - \frac{8\Delta_Z E_B}{\Delta_R^2} \text{Log} \frac{\Delta_Z E_B}{2\Delta_R^2} \bigg) - \Delta_y^2 \bigg( \frac{2}{\Delta_R^2} \text{Log} \frac{\Delta_Z E_B}{2\Delta_R^2} \bigg) \\
&\hspace{1cm} - v_F q \Delta_y \bigg( \frac{8\Delta_Z}{\Delta_R^3} \text{Log} \frac{\Delta_Z E_B}{2\Delta_R^2} \bigg) - \frac{4v_F^2 q^2}{\Delta_R^2} \bigg( \frac{4m^2\alpha^4 - 2\Delta_Z^2}{\Delta_R^2} \bigg) \text{Log} \frac{\Delta_Z E_B}{2\Delta_R^2} + \mathcal{O} (q^2 \Delta_y^2) \Bigg], \\
\Pi_{xy} &= \frac{\nu_F}{8} \Bigg[ \bigg( i\frac{4\Delta_Z^2}{\Delta_R^2} \text{Log} \frac{\Delta_Z E_B}{2\Delta_R^2} \bigg) + \Delta_y^2 \bigg( i\frac{2\Delta_Z^2}{\Delta_R^4} \text{Log} \frac{\Delta_Z E_B}{2\Delta_R^2} \bigg) \\
&\hspace{1cm} + i\frac{v_F q \Delta_y}{E_B} \frac{4\Delta_Z^2}{\Delta_R^3} + i \frac{2v_F^2 q^2}{\Delta_R^2} \frac{\Delta_z (4m^2\alpha^4 - \Delta_Z^2)}{E_B \Delta_R^2} + \mathcal{O} (q^2 \Delta_y^2) \Bigg], \\
\Pi_{yx} &= -\Pi_{xy}, \\
\Pi_{xz} &= \frac{\nu_F}{8} \Bigg[ i\frac{4\Delta_y\Delta_Z}{\Delta_R^2} \bigg( 1 - \frac{E_B \Delta_Z}{\Delta_R^2} \text{Log} \frac{\Delta_Z E_B}{2\Delta_R^2} \bigg) - i \frac{4 v_F q \Delta_Z^2}{\Delta_R^3} \text{Log} \frac{\Delta_Z E_B}{2\Delta_R^2} + \mathcal{O} (q \Delta_y^2) \Bigg], \\
\Pi_{zx} &= -\Pi_{xz}, \\
\Pi_{yz} &= \frac{\nu_F}{8} \Bigg[ \frac{4\Delta_y\Delta_Z}{\Delta_R^2} \bigg( 1 - \frac{E_B \Delta_Z}{\Delta_R^2} \text{Log} \frac{\Delta_Z E_B}{2\Delta_R^2} \bigg) - \frac{4 v_F q \Delta_Z^2}{\Delta_R^3} \text{Log} \frac{\Delta_Z E_B}{2\Delta_R^2} + \mathcal{O} (q \Delta_y^2) \Bigg], \\
\Pi_{zy} &= \Pi_{yz},
\end{split}
\eeq
\ewt
where $E_B$ is the binding energy with the assumption $E_B \ll \Delta_Z, \Delta_R$.
As we see, all the components of the bubble are non-zero at finite $q$ and in the presence of $\Delta_y$, up to $\mathcal{O}(q \Delta_y)$ and $\mathcal{O}(q^2)$. However, we argue that near the lower edge of the continuum, i.e., $\Omega \approx \Delta_Z$, the determinant equation can still be decoupled as in Eq.~\er{det}. 

To begin the discussion, we first consider only $\Delta_y \neq 0$ and set $q=0$. All the charge components of the bubble ($\Pi_{0i}$, $i \in (0...3)$) are zero at $q=0$. The structure of the bubble matrix is now such that the in-plane and the out-of-plane sectors are coupled; this is ensured by $\Pi_{xz}$ and $\Pi_{yz}$ being non-zero. However, it can be argued that at $\Omega \approx \Delta_Z$, $\Pi_{xz}$ and $\Pi_{yz}$ do not affect the determinant equation, $\text{Det}[\hat{M}^{-1}] = 0$ \er{M}, significantly.
Indeed, the determinant equation that is to be solved for $\Omega$ can be written as
\beq
\label{det_tilt}
\bigg( 1 + \frac{U}{2} \Pi_{zz} \bigg) \bigg[ \bigg( 1 + \frac{U}{2} \Pi_{xx} \bigg)^2 + \frac{U^2}{4} \Pi_{xy}^2 \bigg] - \frac{U^3}{4} \Pi_{xy} \Pi_{yz} \Pi_{xz} + \frac{U^2}{4} \bigg( 1 + \frac{U}{2} \Pi_{zz} \bigg) \big( \Pi_{xz}^2 - \Pi_{yz}^2 \big) = 0
\eeq 
Clearly, for $E_B \to 0$, the first term of Eq.~\er{det_tilt} diverges as $\text{log}^2E_B$ while the second term as $\text{log}E_B$: log-divergence in $\Pi_{iz}$, for $i \in (1...3)$, vanishes as $E_B \text{log}E_B$, so their leading contribution is only up to some finite value. This is the reason the third term of Eq.~\er{det_tilt} has no divergence and starts off with a finite value. Hence, compared to the first term, the second term and the third term of Eq.~\er{det_tilt} can be ignored, hence again recovering the decoupled form of the determinant equation to be solved for the collective mode frequency $\Omega$. At weak coupling ($u\ll1$), the finite $\Delta_y$ correction to the collective mode frequency \er{mode}, therefore, mainly comes from the corresponding corrections in $\Pi_{xx}$ and $\Pi_{xy}$.

Now we consider finite $q$ and present the argument on decoupling of \{x, y\}- and \{z\}-sectors once again. We can observe that all the charge components of the bubble, $\Pi_{0i}$ for $i \in (0...3)$, diverges as $q^\alpha \text{log}E_B$. In the spin components, the planar components ($\Pi_{xx}$ and $\Pi_{xy}$) diverge as $q^\alpha/E_B$, while the out-of-plane components ($\Pi_{zz}$, $\Pi_{xz}$ and $\Pi_{yz}$) diverge as $q^\alpha \text{log}E_B$. Clearly, $1/E_B$ divergence is stronger than $\text{log}E_B$, so at finite $q$ the last two terms of Eq.~\er{det_tilt} can be again ignored compared to the first term. 

Finally, at $\Omega \approx \Delta_Z$, the determinant equation \er{det_tilt} can be written in a decoupled form, i.e., keeping only the first term at leading order, and the dispersion of the collective mode \er{mode} can be calculated for tilted magnetic field:
\beq
\label{mode_qx}
\Omega \approx \Delta_Z - \frac{2\Delta_R^2}{\Delta_Z} \text{Exp} \left[ -\frac{\Delta_R^2}{\Delta_Z^2} \frac{1-u}{2u} \left(1 + \frac{\Delta_y^2}{2\Delta_R^2} \right)^{-1} \right] +  v_F q_x \bigg( \frac{\Delta_y}{\Delta_R} \bigg) + v_F^2 q_x^2 \bigg( \frac{4m^2\alpha^4 - \Delta_Z^2}{2\Delta_R^2\Delta_Z} \bigg) + \mathcal{O} (q_x^2 \Delta_y^2).
\eeq
As expected, the leading order $q$-dependent correction to the mode frequency starts off linearly with the amplitude proportional to $\Delta_y$, consistent with the results obtained for the DPV model \er{dis_tilt}.

\subsubsection{$\bq \parallel \hat{y}$}
\label{q_y}
One can similarly calculate the dispersion along $\hat{y}$-direction. However, instead of doing it explicitly, one can directly write down its form knowing that there will not be any linear in $q_y$ (boost) term. Indeed, as discussed in the beginning of this section, $\Delta_y$ breaks mirror symmetry along $\hat{x}$-direction. This naturally explains the leading order $q$-dependent correction to the collective mode frequency be linear in $q_x$. However, since the mirror symmetry along $\hat{y}$ is preserved, the dispersion should be quadratic in $q_y$. From the explicit calculation, we find
\beq
\label{mode_qy}
\Omega \approx \Delta_Z - \frac{2\Delta_R^2}{\Delta_Z} \text{Exp} \left[ -\frac{\Delta_R^2}{\Delta_Z^2} \frac{1-u}{2u} \left(1 + \frac{\Delta_y^2}{2\Delta_R^2} \right)^{-1} \right] + v_F^2 q_y^2 \bigg( \frac{4m^2\alpha^4 - \Delta_Z^2}{2\Delta_R^2\Delta_Z} \bigg) + \mathcal{O} (q_y^2 \Delta_y^2),
\eeq
which is same as that for $\bq \parallel \hat{x}$, except for the linear in $q_x$ term. We note that the term of order $q_{x,y}^2 \Delta_y^2$ is not necessarily going to be the same for $q_x$ and $q_y$.

The coefficient of $q_x^2$ \er{mode_qx} and $q_y^2$ \er{mode_qy} terms in the collective mode has a turning point which has a crucial implication. If for the time being we assume $\Delta_y=0$, then the dispersion along $\hat{x}$ and $\hat{y}$ is quadratic and the coefficient of it is positive when $\Delta_Z < 2m\alpha^2$. 
This is crucial in the sense that if the dispersion along $\hat{z}$ is always monotonic, then the overall dispersion of the collective mode is hyperbolic in nature as long as $\Delta_Z < 2m\alpha^2$. To see this explicitly, we calculate the dispersion along $\hat{z}$-direction in the next section.

\subsubsection{$\bq \parallel \hat{z}$}
\label{q_z}
The general equation of non-interacting polarization bubbles for $\bq$ along $\hat{z}$-direction can be obtained by writing $k_z$ as $k_z+q_z$. Since the coherence factor $f_{ij}$ \er{coh} depends only on in-plane components of $\bk$, the $q$-dependence to the bubble will not arise from here. The only source for $q$-dependence then is the chiral part of the Green's function, $g_s(i\epsilon_m, \bk)$, as defined in Eq.~\er{gr}. 
Near the lower edge of the spin-flip continuum, $\Omega \approx \Delta_Z$, which is possible for interband transitions only, the contribution from charge sector vanishes identically which can be understood from Eq.~\er{coh} by putting $q=0$ and $s=-\bar{s}$. Therefore, only the spin contribution survives, the corresponding coherence factors is same as that in Eq.~\er{coh} at $q=0$. The polarization bubbles can be calculated up to leading singularity as
\bwt
\beq
\label{Pi_qz}
\begin{split}
\Pi_{xx} &= \frac{\nu_F}{8} \Bigg[ \bigg( -4 + \frac{4\Delta_Z^2}{\Delta_R^2} \text{Log} \frac{\Delta_Z E_B}{2\Delta_R^2} \bigg) + \Delta_y^2 \bigg( \frac{2\Delta_Z^2}{\Delta_R^4} \text{Log} \frac{\Delta_Z E_B}{2\Delta_R^2} \bigg) - \frac{2 v_F^2 q^2}{E_B^2} \frac{\Delta_Z^2}{\Delta_R^2} \bigg( 1 - \frac{\Delta_y^2}{2\Delta_R^2} \bigg) \Bigg], \\
\Pi_{yy} &= \Pi_{xx}, \\
\Pi_{zz} &= \frac{\nu_F}{8} \Bigg[ \bigg( -8 - \frac{8\Delta_Z E_B}{\Delta_R^2} \text{Log} \frac{\Delta_Z E_B}{2\Delta_R^2} \bigg) - \Delta_y^2 \bigg( \frac{2}{\Delta_R^2} \text{Log} \frac{\Delta_Z E_B}{2\Delta_R^2} \bigg) - \frac{4 v_F^2 q^2}{\Delta_R^2} \frac{\Delta_Z}{E_B} + \frac{v_F^2 q^2}{\Delta_R^2} \frac{\Delta_y^2}{E_B^2} \Bigg], \\
\Pi_{xy} &= \frac{\nu_F}{8} \Bigg[ i \frac{4\Delta_Z^2}{\Delta_R^2} \text{Log} \frac{\Delta_Z E_B}{2\Delta_R^2} + i \Delta_y^2 \frac{2\Delta_Z^2}{\Delta_R^4} \text{Log} \frac{\Delta_Z E_B}{2\Delta_R^2} - i \frac{2 v_F^2 q^2}{E_B^2} \frac{\Delta_Z^2}{\Delta_R^2} \bigg( 1 - \frac{\Delta_y^2}{2\Delta_R^2} \bigg) \Bigg], \\
\Pi_{yx} &= -\Pi_{xy}, \\
\Pi_{xz} &= \frac{\nu_F}{8} \Bigg[ i\frac{4\Delta_y\Delta_Z}{\Delta_R^4} \bigg( \Delta_R^2 - E_B \Delta_Z \text{Log} \frac{\Delta_Z E_B}{2\Delta_R^2} \bigg) + i \Delta_y \frac{2 v_F^2 q^2 \Delta_Z^2}{\Delta_R^4 E_B} \Bigg], \\
\Pi_{zx} &= -\Pi_{xz}, \\
\Pi_{yz} &= \frac{\nu_F}{8} \Bigg[ \frac{4\Delta_y\Delta_Z}{\Delta_R^4} \bigg( \Delta_R^2 - E_B \Delta_Z \text{Log} \frac{\Delta_Z E_B}{2\Delta_R^2} \bigg) + \Delta_y \frac{2 v_F^2 q^2 \Delta_Z^2}{\Delta_R^4 E_B} \Bigg], \\
\Pi_{zy} &= \Pi_{yz}.
\end{split}
\eeq
\ewt

At small $q$ and $\Delta_y$, the determinant equation \er{det_tilt} (all the charge components are zero, so this equation is valid also at finite $q_z$) can be argued to decouple in the same way as discussed in Sec.~\ref{q_x}.
Assuming $q=0$ first, it can be observed in Eq.~\er{Pi_qz} that $\Pi_{zz}$, $\Pi_{xz}$ and $\Pi_{yz}$ are finite as $E_B \to 0$. On the other hand, all the planar components, such as $\Pi_{xx}$ and $\Pi_{xy}$ diverges logarithmically as $\text{log}E_B$. This is the reason we can ignore $\Pi_{xz}$ and $\Pi_{yz}$ and consider the determinant equation \er{det_tilt} as decoupled.
Now we consider finite $q$. As contrary to that for $q_x$, the $\Pi_{xz}$ and $\Pi_{yz}$ components also have $q^\alpha/E_B$ divergence. However, now the planar components $\Pi_{xx}$ and $\Pi_{xy}$ diverge as $q^\alpha/E_B^2$ which is stronger that $q^\alpha/E_B$. This is the reason we can again consider (also at finite $q_z$) the determinant equation \er{det_tilt} as decoupled.
Using Eq.~\er{plane}, with the form of associated polarization bubbles from Eq.~\er{Pi_qz}, the dispersion of the collective mode \er{mode} along $\hat{z}$-direction can be calculated:
\beq
\label{mode_qz}
\begin{split}
\Omega \approx \Delta_Z - \frac{2\Delta_R^2}{\Delta_Z} \text{Exp} \left[ -\frac{\Delta_R^2}{\Delta_Z^2} \frac{1-u}{2u} \left(1 + \frac{\Delta_y^2}{2\Delta_R^2} \right)^{-1} \right]  - \frac{v_F^2 q_z^2}{\frac{4\Delta_R^2}{\Delta_Z} \text{Exp} \left[ -\frac{\Delta_R^2}{\Delta_Z^2} \frac{1-u}{2u} \left(1 + \frac{\Delta_y^2}{2\Delta_R^2} \right)^{-1} \right]} \left( 1 - \frac{\Delta_y^2}{\Delta_R^2} \right).
\end{split}
\eeq
The dispersion of the mode \er{mode_qz} is monotonic and negative along $\hat{z}$-direction, which, combined with the argument presented above for $\bq$ along $\hat{x}$ \er{mode_qx} and $\hat{y}$ \er{mode_qy} indicate hyperbolicity (assuming $\Delta_y=0$) only when $\Delta_Z < 2m\alpha^2$. If $\Delta_Z > 2m\alpha^2$, then the mode dispersion is parabolic. This is also consistent with our result for the DPV model as obtained in Eq.~\er{dis_tilt}.

We emphasize that one cannot approach to the Silin-Leggett limit here by putting $\Delta_R=0$ due to the same reason mentioned just below Eq.~\er{mode}. Moreover, the approach to the non-interacting limit by putting $u=0$ is also not valid here because the $q$-dependent correction to the binding energy ($E_B$) is obtained perturbatively to the $q=0$ term which required $E_B(q=0) \neq 0$. As we can see from the above equation that since 
$E_B (q=0) \sim e^{-1/u}$, 
the $u=0$ limit makes $E_B(q=0) = 0$, which would violate the perturbation limit above. 

Combining Eqs.~\er{mode_qx}, \er{mode_qy} and \er{mode_qz}, the overall dispersion of the collective mode for the spherical FS model can be written as
\beq
\label{mode_full}
\begin{split}
\Omega &\approx \Delta_z - p_1 - p_2 \delta_{q_y, 0} \delta_{q_z, 0} + \frac{1}{p_2} \frac{\Delta_y^2}{4\Delta_R^2} \left[ \left( v_F q_x + p_2 \frac{2\Delta_R}{\Delta_y} \right)^2 + v_F^2 q_y^2 \right] - v_F^2 q_z^2 \frac{1}{2p_1} \left( 1 - \frac{\Delta_y^2}{\Delta_R^2} \right), \\
& \hspace{0.5cm} \text{where} \,\, p_1 = \frac{2\Delta_R^2}{\Delta_Z} \text{Exp} \left[ -\frac{\Delta_R^2}{\Delta_Z^2} \frac{1-u}{2u} \left(1 + \frac{\Delta_y^2}{2\Delta_R^2} \right)^{-1} \right] \,\, \text{and} \,\, p_2 = \frac{\Delta_y^2 \Delta_Z}{2(4m^2\alpha^4 - \Delta_Z^2)} \\
& \hspace{0.5cm} \text{with} \,\, \delta_{q_i, 0} \,\, \text{as Kronecker delta}. 
\end{split}
\eeq

\bibliography{referenceFile.bib}